\newtheorem{Theorem}{Theorem}
\newtheorem{Lemma}{Lemma}
\newtheorem{Proposition}{Proposition}
\newtheorem{Corollary}{Corollary}
\documentclass[journal]{IEEEtran}
\usepackage{times,amsmath,amssymb,dsfont,graphicx,float, mathrsfs}
\ifCLASSINFOpdf
\else
\fi
\IEEEoverridecommandlockouts

\newcommand{\bbA}{\mathbb{A}}
\newcommand{\bbR}{\mathbb{R}}
\newcommand{\bbI}{\mathbb{I}}

\newcommand{\real}{\mathbb R}
\begin{document}

\title{String Submodular Functions with Curvature Constraints}

\author{Zhenliang~Zhang,~\IEEEmembership{Member,~IEEE},
	Edwin~K.~P.~Chong,~\IEEEmembership{Fellow,~IEEE},
        Ali~Pezeshki,~\IEEEmembership{Member,~IEEE},
        and~William~Moran,~\IEEEmembership{Member,~IEEE}

\thanks{This work was supported in part by AFOSR under Contract FA9550-09-1-0518, and by NSF under Grants CCF-0916314 and CCF-1018472. Part of the results was presented in \cite{2013submodularity}.}
\thanks{Z. Zhang is with Qualcomm Flarion Technology, Bridgewater, NJ 08873 USA. He was with the Department of Electrical and Computer Engineering, Colorado State University, Fort Collins, CO 80523-1373, USA (email: zhenlian@qti.qualcomm.com)
}
\thanks{E. K. P. Chong and A. Pezeshki are with the Department of Electrical and Computer Engineering, Colorado State University, Fort Collins, CO 80523-1373, USA (email: edwin.chong@colostate.edu, ali.pezeshki@colostate.edu)
}
\thanks{W. Moran is with the Department of Electrical and Electronic Engineering, The University of Melbourne, Melbourne, VIC 3010, Australia (email: wmoran@unimelb.edu.au)
}
}

\maketitle

\begin{abstract}
The problem of choosing a string of actions to optimize an objective function that is \emph{string submodular} has been considered in \cite{zhang2012submodularity}. There it is shown that the greedy strategy, consisting of a string of actions that only locally maximizes the step-wise gain in the objective function, achieves at least a $(1-e^{-1})$-approximation to the optimal strategy. This paper improves this approximation by introducing additional constraints on curvature, namely, \emph{total backward curvature, total forward curvature}, and \emph{elemental forward curvature}. We show that if the objective function has {total backward curvature} $\sigma$, then the greedy strategy achieves at least a $\frac{1}{\sigma}(1-e^{-\sigma})$-approximation of the optimal strategy. If the objective function has {total forward curvature} $\epsilon$, then the greedy strategy achieves at least a $(1-\epsilon)$-approximation of the optimal strategy. Moreover, we consider a generalization of the diminishing-return property by defining the {elemental forward curvature}. We also introduce the notion of \emph{string-matroid} and consider the problem of maximizing the objective function subject to a {string-matroid} constraint. We investigate two applications of string submodular functions with curvature constraints: 1) choosing a string of actions to maximize the expected fraction of accomplished tasks; and 2) designing a string of measurement matrices such that the information gain is maximized.

\end{abstract}

\section{Introduction}
\subsection{Background}
We consider the problem of optimally choosing a string of actions over a finite horizon to maximize an objective function. Let $\bbA$ be a set of all possible actions. At each stage $i$, we choose an action $a_i$ from $\bbA$. We use $A=(a_1,a_2,\ldots,a_k)$ to denote a string of actions taken over $k$ consecutive stages, where $a_i\in \bbA$ for $i=1,2,\ldots, k$. We use $\bbA^*$ to denote the set of all possible strings of actions (of arbitrary length, including the empty string). Let $f: \bbA^*\to \bbR$ be an objective function, where $\bbR$ denotes the real numbers. Our goal is to find a string $M\in \bbA^*$, with a length $|M|$ not larger than $K$, to maximize the objective function:
\begin{align}
\begin{array}{l}
\text{maximize  }   f(M) \\
\text{subject to  } M\in\bbA^*, |M|\leq K.
\end{array}
\label{eqn:1}
\end{align}

The solution to \eqref{eqn:1}, which we call the \emph{optimal strategy}, can be found using dynamic programming (see, e.g., \cite{Bertsekas2000}). More specifically, this solution can be expressed with \emph{Bellman's equations}. However, the computational complexity of finding an optimal strategy grows exponentially with respect to the size of $\bbA$ and the length constraint $K$. On the other hand, the greedy strategy, though suboptimal in general, is easy to compute because at each stage, we only have to find an action to maximize the step-wise gain in the objective function. The question we are interested in is: How good is the greedy strategy compared to the optimal strategy in terms of the objective function? This question has attracted widespread interest, which we will review in the next section.

In this paper, we extend the concept of set submodularity in combinatorial optimization to bound the performance of the greedy strategy with respect to that of the optimal strategy. Moreover, we will introduce additional constraints on curvatures, namely, total backward curvature, total forward curvature, and elemental forward curvature, to provide more refined lower bounds on the effectiveness of the greedy strategy relative to the optimal strategy. Therefore, the greedy strategy serves as a good approximation to the optimal strategy.  We will investigate the relationship between the approximation bounds for the greedy strategy and the values of the curvature constraints. These results have many potential applications in closed-loop control problems such as portfolio management (see, e.g., \cite{cover1991universal}), sensor management (see, e.g., \cite{chong2009partially}\cite{li2014suboptimal} \cite{li2014stochastic}), and influence in social networks (see, e.g., \cite{mossel2007submodularity}).

We now provide a simple motivating example in the context of
  sensor management. Suppose that there exists a target of interest
  located in an area deployed with a large number of distributed
  sensors, each of which can detect if the target is located within
  its region of coverage. The goal is to activate sequentially
  individual sensors to maximize the probability of detection of the
  target.  In this context, the action is to activate a sensor at each
  step, and the objective function maximized might be taken to
  be probability of detection, which depends on the sequence of
  sensors activated. Intuitively, without any prior knowledge of
  target location or behavior, and where the sensors have a high
  probability of detection within their individual coverage regions,
  activation of sensors to maximize the total coverage area is a
  suitable surrogate for overall probability of detection. If the
  coverage region of each sensor remains constant  over time, the
  total coverage area will  only depend on the set of sensors activated, and not the order in which they
  are activated, in which case the problem reduces to a special case
  of maximizing a monotone submodular set function subject to a
  knapsack constraint~\cite{tac11}. On the other hand, if the coverage area for each sensor decays
  over time, for instance because of a corrosive environment or
  decaying batteries, the order in which the sensors are activated
  becomes important.  For example, the coverage area for sensor $i$
  might be  given by
  $C_i\exp(-t/t_i)$, where $C_i$ denotes the initial coverage area,
  $t_i$ denotes the mean lifetimes, and $t = 0,1,\ldots$ denotes the
  time index. In these cases, the problem falls within the framework of string submodular
  functions developed in this paper.

\subsection{Related Work}
Submodular set functions play an important role in combinatorial optimization.
Let $X$ be a ground set and $g: 2^X\to \bbR$ be an objective function defined on the power set $2^X$ of $X$. Let $\mathcal I$ be a non-empty collection of subsets of $X$. Suppose that $\mathcal I$ has the \emph{hereditary} and \emph{augmentation} properties: 1. For any $S\subset T\subset X$, $T\in \mathcal I$ implies that $S\in \mathcal I$; 2. For any $S, T \in \mathcal I$, if $T$ has a larger cardinality than $S$, then there exists $j\in T\setminus S$ such that $S\cup \{j\}\in \mathcal I$. Then, we call $(X,\mathcal I)$ a matroid~\cite{tutte1965lectures}. The goal is to find a set in $\mathcal I$ to maximize the objective function:
\begin{align}
\label{eqn:2}
\begin{array}{l}
\text{maximize  }   g(N) \\
\text{subject to  } N\in\mathcal I.
\end{array}
\end{align}
Suppose that $\mathcal I=\{S\subset X: \text{card}(S)\leq k\}$ for a given $k$, where $\text{card}(S)$ denotes the cardinality of $S$. Then, we call $(X,\mathcal I )$ a \emph{uniform} matroid.

The main difference between \eqref{eqn:1} and \eqref{eqn:2} is that the objective function in \eqref{eqn:1} depends on the order of elements in the string~$M$, while the objective function in \eqref{eqn:2} is independent of the order of elements in the set $N$. To further explain the difference, we use $\mathcal P(M)$ to denote a permutation of a string $M$. Note that for $M$ with length $k$, there exist $k!$ permutations.
In \eqref{eqn:1}, suppose that for any $M\in \bbA^*$ we have $f(M)=f(\mathcal P(M))$ for any $\mathcal P$. Then, under these special circumstances, problem \eqref{eqn:1} is equivalent to problem \eqref{eqn:2}. In other words, we can view the second problem as a special case of the first problem. Moreover, there can be repeated identical elements in a string, while a set does not contain identical elements (but we note that this difference can be bridged by allowing the notion of multisets in the formulation of submodular set functions).

Finding the solution to \eqref{eqn:2} is NP-hard---a tractable alternative is to use a greedy algorithm. The greedy algorithm starts with the empty set, and incrementally adds an element to the current solution giving the largest gain in the objective function. Theories for maximizing submodular set functions and their applications have been intensively studied in recent years \cite{nemhauser1978analysis}--\nocite{fisher1978analysis,conforti1984submodular,vondrak2010submodularity,wang2012,streeter2008online,alaei2010maximizing,golovin2011adaptive,buchbinder2012tight,calinescu2011maximizing,chakrabarty2010approximability,vondrak2011submodular,dobzinski2006improved,feige1998threshold,feige2006approximation,feige2010submodular,filmus2012tight,kulik2009maximizing,lee2009non,lee2010submodular,nemhauser1978best,sviridenko2004note,vondrak2008optimal,shamaiah2010greedy,ageev2004pipage}\cite{nesstac}. The main idea is to compare the performance of the greedy algorithm with that of the optimal solution. Suppose that the set objective function $g$ is non-decreasing: $g(A)\leq g(B)$ for all $A\subset B$; and $g(\emptyset)=0$ where $\emptyset$ denotes the empty set. Moreover, suppose that the function has the \emph{diminishing-return property}: For all $A\subset B\subset X$ and $j\in X\setminus B$, we have $g(A\cup \{j\})-g(A)\geq g(B\cup \{j\})-g(B)$. Then, we say that $g$ is a \emph{submodular set} function.
Nemhauser \emph{et al.}~\cite{nemhauser1978analysis} showed that the greedy algorithm achieves at least a $(1-e^{-1})$-approximation for the optimal solution given that $(X, \mathcal I)$ is a uniform matroid and the objective function is submodular. (By this we mean that the ratio of the objective function value of the greedy solution to that of the optimal solution is at least $(1-e^{-1})$.) Fisher \emph{et al.}~\cite{fisher1978analysis} proved that the greedy algorithm provides at least a $1/2$-approximation of the optimal solution for a general matroid. Conforti and Cornu\'{e}jols \cite{conforti1984submodular} showed that if the function $g$ has a total curvature $c$,
where \[c
=\max_{j\in X}\left\{1-\frac{g(X)-g(X\setminus \{j\})}{g(\{j\})-g(\emptyset)}\right\},
\]
then the greedy algorithm achieves at least $\frac{1}{c}(1-e^{-c})$ and $\frac{1}{1+c}$-approximations of the optimal solution given that $(X, \mathcal I)$ is a uniform matroid and a general matroid, respectively. Note that $c\in [0,1]$ for a submodular set function, and if $c=0$, then the greedy algorithm is optimal; if $c=1$, then the result is the same as that in~\cite{nemhauser1978analysis}. Vondr\'{a}k \cite{vondrak2010submodularity} showed that the \emph{continuous greedy algorithm} achieves at least a $\frac{1}{c}(1-e^{-c})$-approximation for any matroid. Wang \emph{et al.}~\cite{wang2012} provided approximation bounds in the case where the function has an elemental curvature $\alpha$, defined as
\[
\alpha
=\max_{S\subset X, i, j \in X, i\neq j}\left\{\frac{g(S\cup \{i,j\})-g(S \cup \{i\})}{g(S\cup \{j\})-g(S)}\right\}.
\] The notion of elemental curvature generalizes the notion of diminishing return. These are powerful results, but are limited in their application to
  optimal control problems that are invariant to the order of
  actions. In most optimal control problems, however, the objective function
  depends crucially on the order of actions and therefore a new
  framework for optimizing objective functions over strings, as
  formulated in~\eqref{eqn:1}, is needed.
  In this paper we do just that in developing string submodularity. We
  will further describe our contributions after reviewing recent results on
  string submodularity.

Some recent papers \cite{zhang2012submodularity}, \cite{streeter2008online}--\nocite{golovin2011adaptive}\cite{alaei2010maximizing} have extended the notion of set
submodularity to problem \eqref{eqn:1}. Streeter and Golovin~\cite{streeter2008online} consider
\eqref{eqn:1} in the context of \emph{online} submodular optimization
(equivalent to our \emph{string} submodular optimization) and  show that if the function $f$ is \emph{prefix} and \emph{postfix} monotone
and $f$ has the diminishing-return property,
then the greedy strategy achieves at least a
$(1-e^{-1})$-approximation of the optimal strategy. Golovin and
Krause~\cite{golovin2011adaptive} introduce the notion of \emph{adaptive}
submodularity for solving stochastic optimization problems under
partial observability. In this Bayesian framework of adaptive
submodularity, the objective function depends on the set of selected
actions; it also depends on a sequence of random outcomes associated
with these actions. At each stage, an action is chosen based on the
previous actions and the random outcomes of these actions. It is
shown in~\cite{golovin2011adaptive} that the greedy strategy achieves at least a
$(1-e^{-1})$ approximation to the expected (calculated with given
prior distributions) objective function of any set of actions. Alaei
and Malekian~\cite{alaei2010maximizing} introduce \emph{sequence} submodularity (also
equivalent to our concept of {string} submodularity) and provide
sufficient conditions for  the  greedy strategy to achieve at least a
$(1-e^{-1})$-approximation to the optimal strategy. Our notion of \emph{string} submodularity and weaker sufficient conditions than those in [14] and [16], under which the greedy
strategy still achieves at least a $(1-e^{-1})$-approximation of the
optimal strategy, were first established in \cite{zhang2012submodularity}.  In contrast to
adaptive submodularity, we take a deterministic approach and
emphasize the importance of the order of actions. The main
contribution of this paper, going  beyond the submodularity bounds
of \cite{zhang2012submodularity}, \cite{streeter2008online}--\nocite{golovin2011adaptive}\cite{alaei2010maximizing}, is the introduction of several notions of
curvature of the string objective function that in turn provide
approximation bounds of the greedy strategy, which are sharper than
$(1-e^{-1})$. We also provide several canonical examples of
applications, different from those
considered in \cite{streeter2008online}--\nocite{golovin2011adaptive}\cite{alaei2010maximizing}, that fall within our framework.

\subsection{Relevance to Control}

While \emph{set} submodular optimization problems are somewhat disconnected with control problems, \emph{string} submodular optimization problems fit right at home in the control literature. Indeed, the problem in \eqref{eqn:1} is, in an unambiguous way, a very general form of an \emph{optimal control} problem. To see this, we rewrite the optimization problem in \eqref{eqn:1} as follows:
\begin{align*}
\mbox{maximize } & f(a_1,\ldots,a_k) \\
\mbox{subject to } & a_i\in\mathbb{A},\ i=1,\ldots,k,\ k\leq K.
\end{align*}
Subject to the monotone assumption, this problem further reduces to
\begin{align*}
\mbox{maximize } & f(a_1,\ldots,a_K) \\
\mbox{subject to } & a_i\in\mathbb{A},\ i=1,\ldots,K.
\end{align*}
This optimization formulation encompasses many optimal control problems that are prevalent in the control literature. Indeed, there seems hardly any way to generalize optimal control problems any further. For example, consider the following optimal control problem:
\begin{align*}
\mbox{minimize } & \sum_{k=1}^K g_k(x_k,u_k) \\
\mbox{subject to } & x_{k+1} = h_k(x_k,u_k),\ k=1,\ldots,K-1,
\end{align*}
where the initial state $x_1$ is given, and $u_k$ is the control
input at time $k$ taking values in some given feasible set. This
formulation is a rather general optimal control problem
involving arbitrary per-stage cost function $g_k$ and state-space
model $x_{k+1} = h_k(x_k,u_k)$. A very special case of this problem
is where the state space is $\real^n$, the feasible input set is
$\real^m$, $g_k(x_k,u_k) = x_k^\top Q x_k + u_k^\top R u_k$ (where $Q$ and $R$ are given symmetric matrices) and $h_k(x_k,u_k) = Ax_k + Bu_k$ (where $A$ and $B$ are appropriately sized matrices). In this case, the problem becomes
\begin{align*}
\mbox{minimize } & \sum_{k=1}^K x_k^\top Q x_k + u_k^\top R u_k \\
\mbox{subject to } & x_{k+1} = Ax_k + Bu_k,\ k=1,\ldots,K,\ x_1\ \mbox{given}.
\end{align*}
This is the familiar finite-horizon, discrete-time \emph{linear quadratic regulator} (LQR) problem, studied in the control literature for decades. In other words, LQR is a very special case of the problem \eqref{eqn:1}.

\subsection{Contributions}
In this paper, we study the problem of maximizing submodular functions defined on strings. We impose additional constraints on curvatures, namely, total backward curvature, total forward curvatures, and elemental forward curvature, which will be rigorously defined in Section II. The notion of total forward and backward curvatures is inspired by the work of  Conforti and Cornu\'{e}jols \cite{conforti1984submodular}. However, the forward and backward algebraic structures are not exposed in the setting of set functions because the objective function defined on sets is independent of the order of elements in a set. The notion of elemental forward curvature is inspired by the work of Wang \emph{et al.} \cite{wang2012}. We have exposed the forward algebraic structure of this elemental curvature in the setting of string functions. Moreover, the result and technical approach in \cite{wang2012} are different from those in this paper. More specifically, the work in~\cite{wang2012} requires the objective function to be a ``set function''; that is, independent of order of elements in the set. In our case, order is a crucial component.

In Section~III, we consider the maximization problem in the case where the strings are chosen from a uniform structure. For this case, our results are summarized as follows. Suppose that the string submodular function $f$ has total backward curvature $\sigma(O)$ with respect to the optimal strategy. Then, the greedy strategy achieves at least a $\frac{1}{\sigma(O)} (1-e^{-\sigma(O)})$-approximation of the optimal strategy. Suppose that the string submodular function $f$ has total forward curvature $\epsilon$. Then, the greedy strategy achieves at least a $(1-\epsilon)$-approximation of the optimal strategy. We also generalize the notion of diminishing return by defining the elemental forward curvature $\eta$. The greedy strategy achieves at least a $1-(1-\frac{1}{K_{\eta}})^K$-approximation, where $K_{\eta}=(1-\eta^K)/(1-\eta)$ if $\eta\neq 1$ and $K_{\eta}=K$ if $\eta= 1$.

In Section~IV, we consider the maximization problem in the case where the strings are chosen from a non-uniform structure by introducing the notion of string-matroid. Our results for this case are as follows. Suppose that the string submodular function $f$ has total backward curvature $\sigma(O)$ with respect to the optimal strategy. Then, the greedy strategy achieves at least a $1/(1+\sigma(O))$-approximation. We also provide approximation bounds for the greedy strategy when the function has total forward curvature and elemental forward curvature.

In Section V, we consider two applications of string submodular functions with curvature constraints: 1) choosing a string of actions to maximize the expected fraction of accomplished tasks; and 2) designing a string of measurement matrices such that the information gain is maximized.

\section{String Submodularity, Curvature, and Strategies}
\subsection{String Submodularity}
We now introduce notation (same to those in \cite{zhang2012submodularity}) to define string submodularity.
Consider a set $\bbA$ of all possible actions. At each stage $i$, we choose an action $a_i$ from $\bbA$. Let $A=(a_1,a_2,\ldots,a_k)$ be a \emph{string} of actions taken over $k$
stages, where $a_i\in \bbA$, $i=1,2,\ldots, k$. Let the set of all possible strings of actions be
\begin{align*}
\bbA^*=\{(a_1,a_2,\ldots,a_k)| & k=0,1,\ldots \text{ and } \\
& a_i\in \bbA,\text{ }  i=1,2\ldots, k\}.
\end{align*}
Note that $k=0$ corresponds to the empty string (no action taken), denoted by $\varnothing$.
For a given string $A=(a_1,a_2,\ldots,a_k)$, we define its \emph{string length} as $k$, denoted $|A|=k$.
If $M=(a_1^m,a_2^m,\ldots, a_{k_1}^m)$ and $N=(a_1^n,a_2^n,\ldots, a_{k_2}^n)$ are two strings in $\bbA^*$, we say $M=N$ if $|M|=|N|$ and $a_i^m=a_i^n$ for each $i=1,2,\ldots, |M|$. Moreover, we define string \emph{concatenation} as follows:
\[
M\oplus N= (a_1^m,a_2^m,\ldots, a_{k_1}^m,a_1^n,a_2^n,\ldots, a_{k_2}^n).
\]
If $M$ and $N$ are two strings in $\bbA^*$, we write $M\preceq N$ if we have
$N=M\oplus L,$
for some $L\in \bbA^*$. In other words, $M$ is a \emph{prefix} of $N$.

A function from strings to real numbers, $f: \bbA^*\to \bbR$, is \emph{string submodular} if
\begin{itemize}
\item[i.] $f$ has the \emph{prefix-monotone} property, i.e.,
\begin{align*}&\quad \forall M, N \in \bbA^*,
& f(M\oplus N)\geq f(M).
\end{align*}
\item[ii.] $f$ has the \emph{diminishing-return} property, i.e.,
\begin{align*}
\nonumber
&\forall M\preceq N \in \bbA^*, \forall a\in \bbA,\\
&f(M\oplus (a))-f(M) \geq f(N\oplus (a))-f(N).
\end{align*}
\end{itemize}

In the rest of the paper, we assume that $f(\varnothing)=0$. Otherwise, we can replace $f$ with the marginalized function $f-f(\varnothing)$. From the prefix-monotone property, we know that $f(M) \geq 0$ for all $M\in \bbA^*$.

We first state an immediate result from the definition of string submodularity.
\begin{Lemma} Suppose that $f$ is string submodular. Then, for any string $N=(n_1,n_2,\ldots, n_{|N|})$, we have
\[
f(N) \leq \sum_{i=1}^{|N|} f((n_i)).
\]
\end{Lemma}
\begin{IEEEproof} We use mathematical induction to prove this lemma. If $|N|=1$, then the result is trivial. Suppose the claim in the lemma holds for any string with length $k$, we wish to prove the claim for any string with length $k+1$. Let $N=(n_1, n_2,\ldots, n_k,n_{k+1})$. By the diminishing return property, we have
\[f((n_{k+1}))-f(\varnothing) \geq f(N)-f((n_1,\ldots,n_k)).\]
Therefore, by the assumption of the induction, we obtain
\[f(N)\leq f((n_{k+1})) +f((n_1,\ldots,n_k)) \leq \sum_{i=1}^{k+1} f((n_i)).\]
This completes the induction proof.
\end{IEEEproof}

Moreover, it is easy to show the following with an induction argument:
\begin{align*}
\nonumber
&\forall M\preceq N \in \bbA^*, \forall L \in \bbA^*,\\
&f(M\oplus L)-f(M) \geq f(N\oplus L)-f(N).
\end{align*}

\subsection{Curvature}

Submodularity in the discrete setting is analogous to concavity in
the continuous setting. Curvature is the degree of this concavity.
In the continuous setting, the degree of concavity is usually captured
using second-order derivatives. In this section, we discuss several measures of curvature to characterize the degree of submodularity, each naturally resulting in a different submodularity performance bound. As we will see below, curvature shares some features of second-order derivatives (or, rather, finite differences).

We define the \emph{total backward curvature} of $f$ by
\begin{align}
\sigma=\max_{a \in \bbA, M\in \bbA^*}\left\{1-\frac{f( (a)\oplus M)-f( M)}{f((a)) -f(\varnothing)}\right\}.
\label{con3}
\end{align}
To see how this notion of curvature is akin to a second-order finite
difference, we first rewrite
\eqref{con3}
 as
\begin{align*}
\sigma &=\max_{a \in \bbA, M\in \bbA^*} \\
&\frac{(f((a))-f(\varnothing))-(f((a)\oplus M)-f(M))}{f((a))-f(\varnothing)}.
\end{align*}
Notice that this is a normalized \emph{difference of differences}
(second-order difference), analogous to a derivative of a derivative
(second-order derivative), so that the total backward curvature is an
upper bound on the second-order difference, over all possible actions
$a$ and strings $M$. Assuming the postfix monotonicity and the
diminishing-return property, all of the differences above are
nonnegative, corresponding to concavity in the string setting.

Similar second-order derivative analogies apply to all the notions of curvature discussed here, including our next definition.
We define the total backward curvature of $f$ with respect to string $M\in \bbA^*$ by
\begin{align}
\sigma(M)=\max_{N\in \bbA^*, 0<|N|\leq K}\left\{1-\frac{f(N\oplus M)-f(M)}{f(N) -f(\varnothing)}\right\},
\label{con4}
\end{align}
where $K$ is the length constraint in \eqref{eqn:1}. Suppose that $f$ is postfix monotone; i.e., $\forall M, N \in \bbA^*,
 f(M\oplus N)\geq f(N).$ Then, we have $\sigma\leq 1$ and $f$ has total backward curvature at most $\sigma$ with respect to any $M\in \bbA^*$; i.e, $\sigma(M) \leq \sigma$ $\forall M\in \bbA^*$. This fact can be shown using a simple derivation: For any $N\in \bbA^*$, we have
\begin{align*}
&f(N\oplus M)-f(M)=\\
&\sum_{i=1}^{|N|} f((n_{i},\ldots,n_{|N|})\oplus M)-f((n_{i+1},\ldots,n_{|N|})\oplus M),
\end{align*}
where $n_i$ represents the $i$th element of $N$. From the definition of total backward curvature and Lemma 1, we obtain
\begin{align*}
f(N\oplus M)-f(M)
&\geq \sum_{i=1}^{|N|} (1-\sigma) f((n_i))\\
&\geq (1-\sigma)f(N),
\end{align*}
which implies that $\sigma(M)\leq \sigma\leq 1$. From the diminishing-return property, it is easy to show that $\sigma \geq 0$.

Symmetrically, we define the \emph{total forward curvature} of $f$ by
\begin{align}
\epsilon=\max_{a \in \bbA, M\in \bbA^*}\left\{1-\frac{f( M\oplus (a))-f( M)}{f((a)) -f(\varnothing)}\right\}.
\end{align}
Moreover, we define the total forward curvature with respect to $M$ by
\begin{align}
\epsilon(M)=\max_{N\in \bbA^*, 0<|N|\leq K}\left\{1-\frac{f(M\oplus N)-f(M)}{f(N) -f(\varnothing)}\right\}.
\end{align}
If $f$ is string submodular and has total forward curvature $\epsilon$, then it has total forward curvature at most $\epsilon$ with respect to any $M\in \bbA^*$; i.e., $\epsilon(M) \leq \epsilon$ $\forall M \in \bbA^*$.
Moreover, for a string submodular function $f$, it is easy to see that for any $M$, we have $\epsilon(M) \leq \epsilon \leq 1$ because of the prefix monotone property and $\epsilon(M) \geq 0$ because of the diminishing-return property.

We define the \emph{elemental forward curvature} of the string submodular function by
\begin{align}
\eta=\max_{a_i, a_j \in \bbA, M\in \bbA^*}{\frac{f(M\oplus (a_i)\oplus (a_j))-f(M\oplus (a_i))}{f(M\oplus (a_j)) -f(M)}}.
\label{con1}
\end{align}
To explain how elemental forward curvature is yet again a form of second-order difference, let us rewrite
\eqref{con1}
 as
\begin{align*}
\eta &=\max_{a_i, a_j \in \bbA, M\in \bbA^*} \\
&{\frac{(f(M\oplus (a_i)\oplus (a_j))-f(M))
-(f(M\oplus (a_i))-f(M))}{f(M\oplus (a_j)) -f(M)}}.
\end{align*}
Again, notice the form of the second-order difference (normalized difference of differences).

In a similar way, we define the \emph{$K$-elemental forward curvature} as follows:
\begin{align}
\nonumber
\hat \eta &=\max_{a_i,a_j\in \bbA,M\in \bbA^*,|M|\leq 2K-2} \\
& \frac{f(M\oplus (a_i) \oplus (a_j))-f(M\oplus (a_i))}{f(M \oplus (a_j))-f(M)}.
\label{curvature}
\end{align}
For a prefix monotone function, we have $\eta\geq 0$, and  the diminishing-return is equivalent to the condition $\eta\leq1$. By the definitions, we know that $\hat \eta \leq \eta$ for all $K$.

The definitions of $\sigma(M)$, $\epsilon(M)$, and $\hat \eta$ depend on the length constraint $K$ of the optimal control problem \eqref{eqn:1}, whereas $\sigma$, $\epsilon$, and $\eta$ are independent of $K$. In other words,
$\sigma$, $\epsilon$, and $\eta$ can be treated as the \emph{universal} upper bounds for $\sigma(M)$, $\epsilon(M)$, and $\hat \eta$, respectively.

\subsection{Strategies}
We will consider the following two strategies.

\emph{1) Optimal strategy}:
Consider the problem \eqref{eqn:1} of finding a string that maximizes $f$ under the constraint that the string length is not larger than $K$. We call a solution of this problem an \emph{optimal strategy} (a term we already have used repeatedly before).
Note that because the function $f$ is prefix monotone, it suffices to just find the optimal strategy subject to the stronger constraint that the string length is equal to $K$. In other words, if there exists an optimal strategy, then there exists one with length $K$.

\emph{2) Greedy strategy}:
A string $G_{k}=(a_1^*,a_2^*,\ldots,a_{k}^*)$ is called \emph{greedy} if $\forall i=1,2,\ldots,k,$
\begin{align*}
a_i^*&=\operatorname*{arg\,max}_{a_i\in \bbA} f((a_1^*,a_2^*,\ldots,a_{i-1}^*,a_i))\\
&\quad-f((a_1^*,a_2^*,\ldots,a_{i-1}^*)).
\end{align*}

Notice that the greedy strategy only maximizes the step-wise~gain in the objective function. In general, the greedy strategy (also called the greedy string) is not an optimal solution to \eqref{eqn:1}. In this paper, we establish theorems which state that the greedy strategy achieves at least a factor of the performance of the optimal strategy, and therefore serves in some sense to \emph{approximate} an optimal strategy.

\section{Uniform Structure}
Let $I$ consist of those elements of $\mathbb A^*$ with maximal length $K$:
$I=\{A\in \bbA^*: |A|\leq K\}.$
We call $I$ a \emph{uniform structure}. Note that the way we define uniform structure is similar to the way we define independent sets associated with uniform matroids. We will investigate the case of non-uniform structure in the next section. Now \eqref{eqn:1} can be rewritten as
\begin{align*}
\begin{array}{l}
\text{maximize  }   f(M) \\
\text{subject to  } M\in I.
\end{array}
\end{align*}

We first consider the relationship between the total curvatures and the approximation bounds for the greedy strategy.

\begin{Theorem} Consider a string submodular function $f$. Let $O$ be a solution to \eqref{eqn:1}. Then, any greedy string $G_K$ satisfies
\begin{itemize}
\item[(i)]
\begin{align*}
f(G_K) &\geq \frac{1}{\sigma(O)}\left(1-\left(1-\frac{\sigma(O)}{K}\right)^{K}\right)f(O)\\
&> \frac{1}{\sigma(O)}(1-e^{-\sigma(O)})f(O),
 \end{align*}
\item[(ii)]  $f(G_K) \geq (1-\max_{i=1,\ldots, K-1} \epsilon(G_i)) f(O).$
\end{itemize}

\label{thm1}
\end{Theorem}
\begin{IEEEproof}
(i)  For any $M\in \bbA^*$ and any $N=(a_1,a_2,\ldots,a_{|N|})\in \bbA^*$, we have
\begin{align*}
&\quad f(M\oplus N)-f(M)\\
&=\sum_{i=1}^{|N|} (f(M\oplus (a_1,\ldots, a_i))
-f(M\oplus (a_1,\ldots, a_{i-1}))) \end{align*}
Therefore, using the prefix monotone property,
there exists an element $a_j\in \bbA$ such that
\begin{align*}
f(M\oplus (a_1,\ldots, a_j))-f(M\oplus (a_1,\ldots, a_{j-1})) \\
 \geq \frac{1}{|N|} (f(M\oplus N)-f(M)).
\end{align*}
Moreover, the diminishing-return property implies that
\begin{align*}
&f(M \oplus (a_j))-f(M) \\
&\geq f(M\oplus (a_1,\ldots, a_j))
-f(M\oplus (a_1,\ldots, a_{j-1}))\\
 &\geq \frac{1}{|N|} (f(M\oplus N)-f(M)).
\end{align*}
Now let us consider the optimization problem \eqref{eqn:1}. Using the property of the greedy strategy and the above inequality (substitute $M=G_{i-1}$ and $N=O$), for each $i=1,2,\ldots,K$ we have
\begin{align*}
&f(G_i)-f(G_{i-1})   \\
& \geq \frac{1}{K} (f(G_{i-1}\oplus O)-f(G_{i-1}))\\
&\geq \frac{1}{K} (f(O)-\sigma(O)f(G_{i-1})).
\end{align*}
Therefore, we have
\begin{align*}
f(G_K)&\geq \frac{1}{K} f(O)+\left(1-\frac{\sigma(O)}{K}\right)f(G_{K-1}) \\
&\geq \frac{1}{K}f(O)\sum_{i=0}^{K-1}  \left(1-\frac{\sigma(O)}{K}\right)^i \\
&=\frac{1}{\sigma(O)}\left(1-\left(1-\frac{\sigma(O)}{K}\right)^{K}\right)f(O).
\end{align*}
Note that \[\frac{1}{\sigma(O)}\left(1-\left(1-\frac{\sigma(O)}{K}\right)^{K}\right)\to \frac{1}{\sigma(O)}(1-e^{-\sigma(O)})\]
from above as $K\to\infty$.
This achieves the desired result.

(ii) Using a similar argument to that in (i), we have
\begin{align*}
&f(G_i)-f(G_{i-1})   \\
& \geq \frac{1}{K} (f(G_{i-1}\oplus O)-f(G_{i-1}))\\
&\geq \frac{1}{K} (f(G_{i-1})+(1-\epsilon(G_{i-1}))f(O)-f(G_{i-1}))\\
&=\frac{1}{K} (1-\epsilon(G_{i-1}))f(O).
\end{align*}
Therefore, by recursion we have
\begin{align*}
f(G_K)&=\sum_{i=1}^K (f(G_i)-f(G_{i-1}) )\\
&\geq\sum_{i=1}^K \frac{1}{K} (1-\epsilon(G_{i-1}))f(O) \\
&\geq  \frac{1}{K} K (1-\max_{i=1,\ldots, K-1} \epsilon(G_i)) f(O) \\
&= (1-\max_{i=1,\ldots, K-1} \epsilon(G_i)) f(O).
\end{align*}

\end{IEEEproof}

Under the framework of maximizing submodular set functions, results with similar form are reported in \cite{conforti1984submodular}. However, the forward and backward algebraic structures are not exposed in \cite{conforti1984submodular} because the total curvature there does not depend on the order of the elements in a set. In the setting of maximizing string submodular functions, the above theorem exposes the roles of forward and backward
algebraic structures in bounding the greedy strategy.
To explain further, let us state the results in a symmetric fashion.
Suppose that the diminishing-return property is stated in a backward way: $f((a)\oplus M)-f(M) \geq f((a)\oplus N)-f(N)$ for all $a\in \bbA$ and $M, N\in \bbA^*$ such that $N=(a_1,\ldots, a_k) \oplus M$. Moreover, a string $\hat G_k=(a_1^*,a_2^*,\ldots,a_{k}^*)$ is called \emph{backward-greedy} if
\begin{align*}
a_i^*&=\operatorname*{arg\,max}_{a_i\in \bbA} f((a_i,a_{i-1}^*,\ldots,a_{2}^*,a_1^*))\\
&\quad-f((a_{i-1}^*,\ldots,a_{1}^*)) \quad \forall i=1,2,\ldots,k.
\end{align*}
Then, we can derive bounds in the same way as Theorem~1, and the results are symmetric.

The results in Theorem 1 implies that for a string submodular function, we have $\sigma(O) \geq 0$. Otherwise, part (i) of Theorem~1 would imply that $f(G_K)\geq f(O)$, which is absurd. Recall that if the function is postfix monotone, then $\sigma(O)\leq \sigma \leq 1$.
 From these facts and part (i) of Theorem~\ref{thm1}, we obtain the following result, also derived in \cite{streeter2008online}.

\begin{Corollary} Suppose that $f$ is string submodular and postfix monotone. Then,
\[
f(G_K) \geq (1-(1-\frac{1}{K})^K)f(O) > (1-e^{-1})f(O).
\]
\label{cor1}
\end{Corollary}

Another immediate result follows from the facts that $\sigma(O)\leq \sigma$ and $\epsilon(G_i)\leq \epsilon$ for all $i$.
\begin{Corollary} Suppose that $f$ is string submodular and postfix monotone. Then,
\begin{itemize}
\item[(i)]
\begin{align*}
f(G_K) &\geq \frac{1}{\sigma}\left(1-\left(1-\frac{\sigma}{K}\right)^{K}\right)f(O)\\
&> \frac{1}{\sigma}(1-e^{-\sigma})f(O),
 \end{align*}
\item[(ii)]  $f(G_K) \geq (1- \epsilon) f(O).$
\end{itemize}
\end{Corollary}

We note that the bounds $\frac{1}{\sigma}(1-e^{-\sigma})$ and $(1-\epsilon)$ are independent of the length constraint $K$. Therefore, the above bounds can be treated as universal lower bounds of the greedy strategy for all possible length constraints.

Next, we use the notion of elemental forward curvature to generalize the diminishing-return property and we investigate the approximation bound using the elemental forward curvature.

\begin{Theorem} Consider a prefix monotone function $f$ with $K$-elemental forward curvature $\hat \eta$ and elemental forward curvature $\eta$. Let $O$ be an optimal solution to \eqref{eqn:1}. Suppose that $f(G_{i}\oplus O) \geq f(O)$ for $i=1,2,\ldots, K-1$. Then, any greedy string $G_K$ satisfies
\begin{align*}
f(G_K)&\geq f(O) \left(1-(1-\frac{1}{K_{\hat \eta}})^{K}\right) \\
& \geq f(O)\left(1-(1-\frac{1}{K_{\eta}})^{K}\right),
\end{align*}
where $K_{\hat \eta}=({1-\hat\eta^{K}})/({1-\hat \eta})$ if $\hat \eta\neq 1$ and $K_{\eta}=K$ if $\hat\eta=1$;
$K_{\eta}= ({1-\eta^{K}})/({1-\eta})$ if $\eta\neq 1$ and $K_{\eta}=K$ if $\eta=1$.
\label{thm2}
\end{Theorem}

\begin{IEEEproof} For any $M,N\in \bbA^*$ such that $|M|\leq K$ and $|N| \leq K$, by the definition of $K$-elemental forward curvature, there exists $a\in \bbA$ such that
\begin{align*}
&f(M\oplus N)-f(M)\\
&=\sum_{i=1}^{|N|} (f(M\oplus (a_1,\ldots, a_i))
-f(M\oplus (a_1,\ldots, a_{i-1})))\\
&\leq \sum_{i=1}^{|N|} \hat \eta^{i-1}( f(M\oplus a_i)-f(M))\\
&\leq (1+\hat\eta+\hat\eta^2+\ldots+\hat\eta^{|N|-1})( f(M\oplus a)-f(M))\\
&=K_{\hat \eta} ( f(M\oplus a)-f(M)).
\end{align*}

Now let us consider the optimization problem \eqref{eqn:1} with length constraint $K$. Using the property of the greedy strategy and the assumptions, we have for $i=1,2,\ldots, K$,
\begin{align*}
&f(G_i)-f(G_{i-1}) \\
&\geq   \frac{1}{K_{\hat \eta}} (f(G_{i-1}\oplus O)-f(G_{i-1}))\\
&\geq\frac{1}{K_{\hat \eta}} (f(O)-f(G_{i-1})).
\end{align*}
Therefore, by recursion, we have
\begin{align*}
f(G_K)&\geq \frac{1}{K_{\hat\eta}} f(O)+(1-\frac{1}{K_{\hat\eta}})f(G_{K-1}) \\
&\geq \frac{1}{K_{\hat\eta}}f(O)\sum_{i=0}^{K-1}  (1-\frac{1}{K_{\hat\eta}})^i \\
&=f(O)\left(1-(1-\frac{1}{K_{\hat\eta}})^{K}\right).
\end{align*}
Because $1-(1-\frac{1}{K_{\hat\eta}})^{K}$ is decreasing as a function of $\hat \eta$ and $\hat \eta \leq \eta$ by definition, we obtain
\begin{align*}
 f(O) \left(1-(1-\frac{1}{K_{\hat \eta}})^{K}\right) \geq f(O)\left(1-(1-\frac{1}{K_{\eta}})^{K}\right).
\end{align*}
\end{IEEEproof}

Recall that $\hat\eta$ depends on the length constraint $K$, whereas $\eta$ does not. Therefore, the lower bound using $K_{\eta}$ can be treated as a universal lower bound of the greedy strategy.

Suppose that $f$ is string submodular. Then, we have $\eta \leq 1$. Because $1-(1-\frac{1}{K_{\eta}})^{K}$ is decreasing as a function of $\eta$, we obtain the following result, which is reported in \cite{zhang2012submodularity}.

\begin{Corollary} Consider a string submodular function $f$. Let $O$ be a solution to \eqref{eqn:1}. Suppose that $f(G_{i}\oplus O) \geq f(O)$ for $i=1,2,\ldots, K-1$. Then, any greedy string $G_K$ satisfies
\[
f(G_K)\geq (1-(1-\frac{1}{K})^K)f(O) > (1-e^{-1})f(O).
\]
\end{Corollary}

The second inequality in the above corollary is given by the fact that $1-(1-\frac{1}{K})^K \to 1-e^{-1}$ from above, as $K$ goes to infinity.
Next we combine the results in Theorems~\ref{thm1} and \ref{thm2} to yield the following result.

\begin{Proposition} Consider a prefix monotone function $f$ with elemental forward curvature $\eta$ and $K$-elemental forward curvature $\hat\eta$. Let $O$ be a solution to \eqref{eqn:1}. Then, any greedy string $G_K$ satisfies
\begin{itemize}
\vspace{0.04in}
\item[(i)] \begin{align*}
f(G_K) &\geq \frac{1}{\sigma(O)}\left(1-\left(1-\frac{\sigma(O)}{K_{ \hat \eta}}\right)^{K}\right)f(O)\\
& \geq  \frac{1}{\sigma(O)}\left(1-\left(1-\frac{\sigma(O)}{K_{ \eta}}\right)^{K}\right)f(O),
 \end{align*}
\vspace{0.04in}
\item[(ii)] \begin{align*}
 f(G_K)& \geq (1-\max_{i=1,\ldots, K-1} \epsilon(G_i)) \frac{K}{K_{ \hat \eta}}f(O) \\
 & \geq(1-\max_{i=1,\ldots, K-1} \epsilon(G_i)) \frac{K}{K_{  \eta}}f(O).
 \end{align*}
\end{itemize}
\end{Proposition}

The proof is given in Appendix A. We note that the condition in Theorem~\ref{thm2}, $f(G_{i}\oplus O) \geq f(O)$ for $i=1,\ldots, K-1$, is essentially captured by $\sigma(O)$. In other words, even if the condition $f(G_{i}\oplus O) \geq f(O)$ is violated, we can still provide approximation bound using $\sigma(O)$, which is larger than 1 in this case.

\section{Non-uniform Structure}

In the last section, we considered the case where $I$ is a uniform structure. In this section, we consider the case of non-uniform structures.

We first need the following definition. Let $M=(m_1,\ldots,m_{|M|})$ and $N=(n_1,\ldots,n_{|N|})$ be two strings in $\bbA^*$. We write $M\prec N$ if there exists a sequence of strings $L_i\in \bbA^*$ such that
\begin{align*}
N&=L_1 \oplus (m_1, \ldots, m_{i_1}) \oplus L_2 \oplus (m_{i_1+1},\ldots, m_{i_2})\oplus \mbox{}\\
 &\ldots \oplus (m_{i_{k-1}+1},\ldots, m_{|M|})\oplus L_{k+1}.
\end{align*}
In other words, we can remove some elements in $N$ to get $M$. Note that $\prec$ is a weaker notion of dominance than $\preceq$ defined earlier in Section II. In other words, $M\preceq N$ implies that $M\prec N$ but the converse is not necessarily true.

Now we state the definition of a non-uniform structure, analogous to the definition of independent sets in matroid theory.
A subset $I$ of $\bbA^*$ is called a \emph{non-uniform structure} if it satisfies the following conditions:
\begin{itemize}
\item[1.] $I$ is non-empty;
\item[2.] \emph{Hereditary}: $\forall M\in I$, $N\prec M$ implies that $N\in I$;
\item[3.] \emph{Augmentation}: $\forall M, N \in I$ and $|M|<|N|$, there exists an element $x\in \bbA$ in the string $N$ such that $M\oplus (x)\in I$.
\end{itemize}
By analogy with the definition of a matroid, we call the pair $(\bbA,I)$ a \emph{string-matroid}.
We assume that there exists $K$ such that for all $M\in I$ we have $|M|\leq K$ and there exists a $N \in I$ such that $|N|=K$. We call such a string $N$ a \emph{maximal} string. We are interested in the following optimization problem:
\begin{align}
\label{eqn:n}
\begin{array}{l}
\text{maximize  }   f(N) \\
\text{subject to  } N\in I.
\end{array}
\end{align}
Note that if the function is prefix monotone, then the maximum of the function subject to a string-matroid constraint is achieved at a maximal string in the matroid.
The greedy strategy $G_k=(a_1^*,\ldots, a_k^*)$ in this case is given by
\begin{align*}
a_i^*&=\operatorname*{arg\,max}_{a_i\in \bbA \text{ and } (a_1^*,\ldots,a_{i-1}^*,a_i )\in I} f((a_1^*,a_2^*,\ldots,a_{i-1}^*,a_i))\\
&\quad-f((a_1^*,a_2^*,\ldots,a_{i-1}^*)) \quad \forall i=1,2,\ldots, k.
\end{align*}
Compared with \eqref{eqn:1}, at each stage $i$, instead of choosing $a_i$ arbitrarily in $\bbA$ to maximize the step-wise gain in the objective function, we also have to choose the action $a_i$ such that the concatenated string $(a_1^*,\ldots,a_{i-1}^*,a_i )$ is an element of the non-uniform structure $I$. We first establish the following theorem.

\begin{Theorem} For any $N\in I$, there exists a permutation of $N$, denoted by $\mathcal P(N)=(\hat n_1,\hat n_2,\ldots, \hat n_{|N|})$, such that for $i=1,2,\ldots, |N|$ we have  \[
f(G_{i-1}\oplus (\hat n_i))-f(G_{i-1}) \leq f(G_i)-f(G_{i-1}).
\]
\label{thm3}
\end{Theorem}
\begin{IEEEproof}
We prove this claim by induction on $i=|N|,|N|-1,\ldots,1$ (in descending order). If $i=|N|$, considering $G_{|N|-1}$ and $N$, we know from the String-Matroid Axiom 3 that there exists an element of $N$, denoted by $\hat n_{|N|}$ (we can always permute this element to the end of the string with a certain permutation), such that $G_{|N|-1}\oplus (\hat n_{|N|})\in  I$. Moreover, we know that the greedy way of selecting $a_{|N|}^*$ gives the largest gain in the objective function. Therefore, we obtain
\[f(G_{|N|-1}\oplus (\hat n_i))-f(G_{|N|-1}) \leq f(G_{|N|})-f(G_{|N|-1}).\]

Now let us assume that the claim holds for all $i> i_0$ and the corresponding elements are $\{\hat n_{i_0+1},\ldots, \hat n_{|N|}\}$. Next we show that the claim is true for $i=i_0$. Let $\hat N_{i_0}$ be the string after we remove the elements in $\{\hat n_{i_0+1},\ldots, \hat n_{|N|}\}$ from the original string $N$. We know from Axiom 2 that $\hat N_{i_0}\in I$ and that $|G_{i_0-1}|<|\hat N_{i_0}|$, therefore, there exists an element from $\hat N_{i_0}$, denoted by $\hat n_{i_0}$, such that $G_{i_0-1}\oplus (\hat n_{i_0})\in  I$. Using the property of the greedy strategy, we obtain
\[f(G_{i_0-1}\oplus (\hat n_{i_0}))-f(G_{i_0-1}) \leq f(G_{i_0})-f(G_{i_0-1}).\] This concludes the induction proof.
\end{IEEEproof}

Next we investigate the approximation bounds for the greedy strategy using the total curvatures.
\begin{Theorem} Let $O$ be an optimal strategy for \eqref{eqn:n}. Suppose that $f$ is a string submodular function. Then, a greedy strategy $G_K$ satisfies
\begin{itemize}
\vspace{0.02in}
\item[(i)]
$f(G_K) \geq \frac{1}{1+\sigma(O)} f(O),$
\vspace{0.02in}
\item[(ii)] $f(G_K) \geq (1-\epsilon(G_K)) f(O)$.
\end{itemize}
\label{thm4}
\end{Theorem}
\begin{IEEEproof}
(i) By the definition of the total backward curvature, we know that
\[
f(G_K\oplus O)-f(O) \geq (1-\sigma(O))f(G_K).
\]
Therefore, we have
\begin{align*}
&f(O) \\
& \leq f(G_K\oplus O) -(1-\sigma(O))f(G_K)\\
& =f(G_K)-(1-\sigma(O))f(G_K) +f(G_K\oplus O) -f(G_K).
\end{align*}
Let $O=(o_1,o_2,\ldots, o_{K})$. By the diminishing-return property, we have
\begin{align*}
&f(G_K\oplus O) -f(G_K) \\
&=\sum_{i=1}^{K} (f(G_K \oplus (o_1,\ldots, o_i))-f(G_K \oplus (o_1,\ldots, o_{i-1}))) \\
&\leq \sum_{i=1}^{K} (f(G_K \oplus (o_i))-f(G_K)).
\end{align*}
By Theorem~3, we know that there exists a permutation of $O$: $\mathcal P(O)=(\hat o_1,\hat o_2,\ldots, \hat o_{|O|})$ such that
\[
f(G_{i-1}\oplus (\hat o_i))-f(G_{i-1}) \leq f(G_i)-f(G_{i-1}),
\]
for $i=1,2,\ldots, K$.
Therefore, by the diminishing-return property again,
\begin{align*}
& \quad  \sum_{i=1}^{K} (f(G_K \oplus (o_i))-f(G_K))\\
&\leq\sum_{i=1}^{K} ( f(G_{i-1}\oplus (\hat o_i))-f(G_{i-1})) \\
&\leq \sum_{i=1}^{K} (f(G_i)-f(G_{i-1})) \\
&= f(G_K).
\end{align*}
From the above equations,
\begin{align*}
f(O)& \leq f(G_K)+f(G_K)-(1-\sigma(O))f(G_K)\\
&=(1+\sigma(O))f(G_K),
\end{align*}
and this achieves the desired result.

(ii) From the definition of total forward curvature, we have
\[
f(G_K\oplus O)-f(G_K) \geq (1-\epsilon(G_K)) f(O).
\]
From the proof of part (i), we also know that $f(G_K\oplus O)-f(G_K) \leq f(G_K)$.
Therefore, we have $f(G_K) \geq (1-\epsilon(G_K)) f(O).$
\end{IEEEproof}

The inequality in (i) above is a generalization of a result on maximizing submodular set functions with a general matroid constraint \cite{fisher1978analysis}.  The submodular set counterpart involves total curvature, whereas the string version involves total \emph{backward} curvature. Note that if $f$ is postfix monotone, then $\sigma(O)\leq \sigma \leq  1$. We now state an immediate corollary of Theorem~\ref{thm4}.
\begin{Corollary}
Suppose that $f$ is string submodular and postfix monotone. Then, the greedy strategy achieves at least a $1/2$-approximation of the optimal strategy.

\end{Corollary}

Another immediate result follows from the facts that $\sigma(O)\leq \sigma$ and $\epsilon(G_K)\leq \epsilon$.
\begin{Corollary}
Suppose that $f$ is string submodular and postfix monotone. Then, we have
\begin{itemize}
\vspace{0.02in}
\item[(i)]
$f(G_K) \geq \frac{1}{1+\sigma} f(O),$
\vspace{0.02in}
\item[(ii)] $f(G_K) \geq (1-\epsilon) f(O)$.
\end{itemize}
\end{Corollary}

Next we generalize the diminishing-return property using the elemental forward curvature.
\begin{Theorem} Suppose that $f$ is a prefix monotone function with elemental forward curvature $\eta$ and $K$-elemental forward curvature $\hat \eta$. Suppose that $f(G_K\oplus O)\geq f(O)$. If $\hat\eta \leq 1$, then
\[
f(G_K) \geq \frac{1}{1+\hat\eta} f(O) \geq \frac{1}{1+\eta} f(O).
\]
If $\hat\eta >1$, then
\[
f(G_K) \geq \frac{1}{1+\hat\eta^{2K-1}} f(O) \geq \frac{1}{1+\eta^{2K-1}} f(O).
\]
\label{thm5}
\end{Theorem}
\begin{IEEEproof} Let $O=(o_1,o_2,\ldots, o_K)$.
From the definition of $K$-elemental forward curvature,
\begin{align*}
&f(G_K\oplus O)  -f(G_K) \\
&=\sum_{i=1}^{K} (f(G_K\oplus (o_1,\ldots,o_i)) -f(G_K\oplus (o_1,\ldots, o_{i-1})))\\
& \leq \sum_{i=1}^{K} (f(G_{K-1}\oplus (o_i))-f(G_{K-1}))\hat\eta^i \\
&\leq \begin{cases} \hat \eta \sum_{i=1}^{K} (f(G_{K-1}\oplus (o_i))-f(G_{K-1})), & \mbox{if } \hat\eta\leq 1 \\ \hat\eta^{K} \sum_{i=1}^{K} (f(G_{K-1}\oplus (o_i))-f(G_{K-1})), & \mbox{if } \hat\eta> 1. \end{cases}
\end{align*}

From Theorem~3, there exists a permutation $\mathcal P$ of $O$: $\mathcal P(O)=(\hat o_1,\ldots, \hat o_K)$, such that
\[
f(G_{i-1}\oplus (\hat o_i))-f(G_{i-1}) \leq f(G_i)-f(G_{i-1}),
\]
for $i=1,\ldots, K$.
Moreover, by the definition of $K$-elemental forward curvature,
\begin{align*}
 &\quad \sum_{i=1}^{K} (f(G_{K-1}\oplus (o_i))-f(G_{K-1}))   \\
 &=\sum_{i=1}^{K} (f(G_{K-1}\oplus (\hat o_i))-f(G_{K-1}))   \\
 &\leq  \sum_{i=1}^{K} \hat\eta^{K-i} (f(G_{i-1}\oplus (\hat o_i))-f(G_{i-1}))  \\
&\leq \begin{cases}  \sum_{i=1}^{K} (f(G_{i-1}\oplus (\hat o_i))-f(G_{i-1})), & \mbox{if } \hat\eta\leq 1 \\ \hat\eta^{K-1} \sum_{i=1}^{K} (f(G_{i-1}\oplus (\hat o_i))-f(G_{i-1})), & \mbox{if } \hat\eta> 1. \end{cases}\\
&\leq \begin{cases}  f(G_K), & \mbox{if } \hat\eta\leq 1 \\ \hat\eta^{K-1}f(G_K), & \mbox{if } \hat\eta> 1. \end{cases}
\end{align*}
Therefore, we have
\begin{align*}
f(O)\leq \begin{cases} (1+\hat\eta) f(G_K), & \mbox{if } \hat\eta\leq 1 \\ (1+\hat\eta^{2K-1})f(G_K), & \mbox{if } \hat\eta> 1. \end{cases}
\end{align*}
Since $\hat \eta \leq \eta$ and $\frac{1}{1+\hat\eta}$ and $\frac{1}{1+\hat\eta^{2K-1}}$ are monotone decreasing functions of $\hat \eta$, we obtain the desired results.
\end{IEEEproof}

This result is similar in form to that in \cite{wang2012}. However, the second bound in Theorem~\ref{thm5} is different from that in \cite{wang2012}. This is because the proof in \cite{wang2012} uses the fact that the value of a set function at a set is independent of the order of elements in the set, whereas this is not the case for a string.
Recall that the elemental forward curvature for a string submodular function is not larger than 1. We obtain the following result.

\begin{Corollary} Suppose that $f$ is a string submodular function and $f(G_K\oplus O)\geq f(O)$. Then, the greedy strategy achieves at least a $1/2$-approximation of the optimal strategy.
\end{Corollary}

Now we combine the results for total and elemental curvatures to get the following.

\begin{Proposition} Suppose that $f$ is a prefix monotone function with $K$-elemental forward curvature $\hat\eta$ and elemental forward curvature $\eta$. Then, a greedy strategy $G_K$ satisfies
\begin{itemize}
\vspace{0.04in}
\item[(i)] $f(G_K) \geq\frac{1}{\sigma(O)+ h(\hat\eta)} f(O) \geq \frac{1}{\sigma(O)+ h(\eta)} f(O) $,
\vspace{0.04in}
\item[(ii)] $f(G_K) \geq \frac{1-\epsilon(G_K)}{ h(\hat\eta)}f(O)\geq \frac{1-\epsilon(G_K)}{ h(\eta)}f(O)$,
\end{itemize}
where $h(\hat\eta)=\hat\eta$ and $h(\eta)=\eta$ if $\hat\eta \leq 1$; $h(\hat\eta)= \hat\eta^{2K-1}$ and $h(\eta)= \eta^{2K-1}$ if $\hat\eta >1$.
\end{Proposition}

The proof is given in Appendix B. From these results, we know that when $f$ is string submodular, $\hat \eta \in [0,1]$ and we must have $
\sigma(O)+ \hat \eta \geq 1$ and
$\epsilon(G_K)+\hat\eta \geq 1.$
From Theorems~\ref{thm1}, \ref{thm2}, \ref{thm4}, and \ref{thm5}, we see that the performance of the greedy strategy relative to the optimal improves as the total forward/backward curvature or the elemental forward curvature decreases to $0$. On the other hand, the inequalities above indicate that this performance improvement with forward and elemental curvature constraints cannot become arbitrarily good simultaneously.
When equality in either case holds, the greedy strategy is optimal. A special case for this scenario is when the objective function is \emph{string-linear}: $f(M\oplus N)=f(M)+f(N)$ for all $M,N \in \bbA^*$, i.e., $\eta =1$ and $\sigma=\epsilon=0$. Recall that $0\leq \sigma(O)\leq \sigma$, $0\leq \epsilon(G_K)\leq \epsilon$, and $0 \leq \hat \eta\leq \eta$. Therefore, we have $\sigma(O)=\epsilon(G_k)=0$ and $\hat \eta=1$.

\emph{Remark}: The above proposition and the discussions afterward easily generalize to the framework of submodular set functions.

\section{Applications}
In this section, we investigate two applications of string submodular functions with curvature constraints. We note that explicitly computing all the curvatures defined
    in the previous sections might not always be feasible. However, as
    we shall see later in this section, in some canonical example
    applications, we can either compute the curvature explicitly or
    provide tight bounds for the curvature, which in turn bound the
    performance of the greedy strategy.
\subsection{Strategies for Accomplishing Tasks}
Consider an objective function of the following form:
\begin{align}
\label{app1}
f((a_1,\ldots, a_k))=\frac{1}{n} \sum_{i=1}^n \left( 1-\prod_{j=1}^k (1-p_i^j(a_j))\right).
\end{align}
We can interpret this objective function as follows. We have $n$ subtasks, and by choosing action $a_j$ at stage $j$ there is a probability $p_i^j(a_j)$ of accomplishing the $i$th subtask. Therefore, the objective function is the expected fraction of subtasks that are accomplished after performing $(a_1,\ldots, a_k)$. A special case of this problem has been studied in \cite{streeter2008online}, where $p_i^j(a_j)$ only depends on $a_j$ the time $t_j$ invested in stage $j$. It is shown there that if $p_i^j(a_j)$ is a non-decreasing function of $t_j$ for all $i$ and $a_j$, then the greedy strategy achieves at least $(1-e^{-1})$-approximation to the optimal strategy. We will reinvestigate the general case \eqref{app1} using the aforementioned notions of curvature and string-matroid. Note that if $p_i^j$ is independent of $j$ for all $i$; i.e., the probability of accomplishing the $i$th subtask by choosing an action does not depend on the stage at which the action is chosen, then it is obvious that the objective function does not depend on the order of actions. In this special case, the objective function is a submodular set function and therefore the greedy strategy achieves at least a $(1-e^{-1})$-approximation of the optimal strategy. Moreover, this special case is closely related to several previously studied problems, such as min-sum set cover \cite{feige2004approximating}, pipelined set cover \cite{munagala2005pipelined}, social network influence \cite{kempe2005influential}, and coverage-aware self scheduling in sensor networks \cite{lu2003coverage}. In this paper, we generalize the special case to the situation where $p_i^j$ depends on $j$. Applications of this generalization include designing campaign strategy for political voting and scheduling problems in control literature~\cite{tacScheduling}. Without loss of generality, we will consider the special case where $n=1$ (our analysis easily generalizes to arbitrary $n$). In this case, we have
\[
f((a_1,\ldots, a_k))=  1-\prod_{j=1}^k (1-p^j(a_j)).
\]
For each $a\in \bbA$, we assume that $p^j(a)$ takes values in $[L(a),U(a)]$, where $0<L(a)<U(a)<1$. Moreover, let
\[
c(a)=\frac{1-U(a)}{1-L(a)}.
\]
Obviously, $c(a) \in (0,1)$. The prefix monotone property is easy to check:
For any $M, N\in \bbA^*$, the statement that $f(M\oplus N)\geq f(M)$ is obviously true.

\subsubsection{Uniform Structure}
We first consider the maximization problem under the uniform structure constraint. We have the following results.
\begin{Theorem}\label{thm:app1}
Let $\hat U=\max_{a\in \bbA} U(a)$, $\hat L=\min_{a\in \bbA} L(a)$, and $c=\min_{a\in \bbA} c(a)$. Suppose that $\hat L^{-1} - \hat U^{-1} \leq 1$. Then, we have
\begin{itemize}
\item[i)]
\begin{align*}
f(G_K) \geq \frac{1}{\bar \sigma} \left(1-\left(1-\frac{\bar \sigma}{K}\right)^K\right) f(O),
\end{align*}
where $\bar \sigma =1- \min_{K \leq k < 2K}\frac{(1-\hat U)^{k} -(1-\hat L)^{k+1}}{\hat L}$.

\item[ii)]  \[f(G_K) \geq \frac{(1-\hat U)^{2K-2}\hat L}{\hat U} f(O).\]

\item[iii)] if $p^1(a_1^*) \geq 1- c^K$, where $a_1^*$ represent the greedy action at stage 1, then\begin{align*}
f(G_K)&\geq f(O) \left(1-(1-\frac{1}{K_{ \eta}})^{K}\right) \geq \left(1-(1-\frac{1}{K_{ \bar \eta}})^{K}\right),
\end{align*}
where $K_{\eta}= ({1-\eta^{K}})/({1-\eta})$ and $\eta =\max_{a_{i}, a_{j}} \frac{(1-p^{i}(a_{i}))p^{j}(a_j) }{p^i(a_j)}$; $K_{\bar\eta}= ({1-\bar\eta^{K}})/({1-\bar\eta})$ and $\bar \eta = \frac{(1-\hat L)\hat U}{\hat L}$.
\end{itemize}
\end{Theorem}

\begin{IEEEproof}
i) The elemental forward curvature in this case is
\[
\eta =\max_{a_{i}, a_{j}} \frac{(1-p^{i}(a_{i}))p^{j}(a_j) }{p^i(a_j)}.
\]
 Then, from the definitions, we have \[
\eta\leq \frac{(1-\hat L)\hat U}{\hat L}.
\]
Note that the function is submodular if and only if $\eta\leq 1$. From the above equation, we conclude that $f$ is submodular if
\[ \frac{(1-\hat L)\hat U}{\hat L}\leq 1.\]
Therefore, a sufficient condition for $f$ to be a string submodular function is
\[
\hat L^{-1}-\hat U^{-1} \leq 1.
\]

To apply Theorem~\ref{thm1}, instead of calculating the total backward curvature with respect to the optimal strategy, we calculate the total backward curvature for $K\leq|M|< 2K$:
\begin{align}
\hat\sigma&=\max_{a \in \bbA, K\leq|M|< 2K}\left\{1-\frac{f(( a)\oplus M)-f( M)}{f((a)) -f(\varnothing)}\right\}\\
&=1-\min_{a \in \bbA, K\leq|M|< 2K}\left\{\frac{f( (a)\oplus M)-f( M)}{f( (a)) -f(\varnothing)}\right\}.
\label{con10}
\end{align}
We have
\begin{align*}
&\quad\frac{f( (a)\oplus M)-f( M)}{f( (a)) -f(\varnothing)}\\
&=\frac{\prod_{j=1}^{|M|}(1-p^j(a_j))-(1-p^1(a))\prod_{j=1}^{|M|}(1-p^{j+1}(a_{j}))}{p^1(a)}.
\end{align*}
We then provide an upper bound for the total backward curvature for all possible combination of $p^j$. The minimum of the above term is achieved at $p^j(a_j)=\hat U$ and $p^{j+1}(a_{j})=\hat L$:
\begin{align*}
&\min_{a \in \bbA, K\leq|M|< 2K}\left\{\frac{f( (a)\oplus M)-f( M)}{f( (a)) -f(\varnothing)}\right\} \\
&\geq \min_{a \in \bbA, K\leq k< 2K} \frac{(1-\hat U)^k-(1-p^1(a))(1-\hat L)^k }{p^1(a)}\\
&\geq\min_{K \leq k < 2K}\frac{(1-\hat U)^{k} -(1-\hat L)^{k+1}}{\hat L} := 1-\bar \sigma.
\end{align*}
Moreover, it is easy to verify that $\sigma(O) \leq \bar \sigma$. Therefore, we can substitute the above upper bound of $\bar \sigma$ to Theorem 1 to derive a lower bound for the approximation of the greedy strategy.

ii) Instead of calculating the total forward curvature with respect to the greedy strategy $G_i$, we calculate
\begin{align}
\hat\epsilon_i&=\max_{a \in \bbA, i\leq|M|< i+K}\left\{1-\frac{f(M \oplus (a))-f( M)}{f((a)) -f(\varnothing)}\right\}\\
&=1-\min_{a \in \bbA, i\leq|M|< i+K}\left\{\frac{f( M\oplus (a))-f( M)}{f(( a)) -f(\varnothing)}\right\}\\
&=1-\min_{a\in \bbA, i\leq|M|< i+K}\frac{\prod_{j=1}^{|M|}(1-p^j(a_j))p^{|M|+1}(a)}{p^1(a)}\\
&\leq 1-(1-\hat U)^{i+K-1}\hat L/\hat U.
\end{align}
It is easy to show that $\epsilon(G_i) \leq \hat  \epsilon_i$. Moreover,
\begin{align*}
\max_{i=1,\ldots, K-1} \epsilon(G_i)\leq \max_{i=1,\ldots, K-1} \hat \epsilon_i \leq  1-(1-\hat U)^{2K-2}\hat L/\hat U.
\end{align*}
We can substitute this upper bound in Theorem~\ref{thm1} and get a lower bound for the approximation of the optimal strategy that the greedy strategy is guaranteed to achieve.

iii) We will use the results in Theorem~\ref{thm2}, which requires the assumption that $f(G_i\oplus O) \leq f(O)$ for $i=1,\ldots, K-1$, which can be written as (assuming $G_i =(a_1^*,\ldots, a_i^*)$ and $O=(o_1,\ldots, o_K)$)
\begin{align}
\prod_{j=1}^K (1-p^j(o_j)) \geq \prod_{t=1}^i (1- p^t(a_t^*)) \prod_{j=1}^{K} (1-p^{j+i}(o_j)).
\label{sf}
\end{align}
We know that
\[
\prod_{j=1}^K (1-p^j(o_j)) \geq \prod_{j=1}^K (1- U(o_j))
\]
and
\begin{align*}
&\prod_{t=1}^i (1- p^t(a_t^*)) \prod_{j=1}^{K} (1-p^{j+i}(o_j))\\
\leq & \prod_{j=1}^K (1- L(o_j)) (1-p^1(a_1^*)).
\end{align*}
Therefore, a sufficient condition for \eqref{sf} is
\[
1-p^1(a_1^*)  \leq \frac{\prod_{j=1}^K (1- U(o_j))}{\prod_{j=1}^K (1- L(o_j))}=\prod_{j=1}^K c(o_j).
\]
This inequality holds because of the assumption that $
p^1(a_1^*) \geq 1- c^K.$ The bound simply follows from the definition of elemental curvature.
\end{IEEEproof}
We note that with additional side information, we can improve the bounds in Theorem~\ref{thm:app1}. For example, suppose that $(1-\hat U)^k/(1-\hat L)^k \geq 1-\hat L$ for all $K\leq k<2K$. Then, we have
\begin{align*}
\bar \sigma & = 1-\min_{K \leq k < 2K}\frac{(1-\hat U)^{k} -(1-\hat L)^{k+1}}{\hat L}\\
&=1-\frac{(1-\hat U)^{2K-1} -(1-\hat L)^{2K}}{\hat L}.
\end{align*}
Furthermore, recall that $\sigma\leq 1$ if the function is postfix monotone. In this case, the value of $\bar\sigma$ in part i) of Theorem~\ref{thm:app1} can be written as $\bar \sigma = \min \left\{1- \min_{K \leq k < 2K}\frac{(1-\hat U)^{k} -(1-\hat L)^{k+1}}{\hat L},1\right\}$.

\subsubsection{Non-uniform Structure}
The calculation for the case of non-uniform structure uses a similar analysis. We have the following results.

\begin{Theorem}
Let $\hat U=\max_{a\in \bbA} U(a)$, $\hat L=\min_{a\in \bbA} L(a)$, and $c=\min_{a\in \bbA} c(a)$. Suppose that $\hat L^{-1} - \hat U^{-1} \leq 1$. Then, we have
\begin{itemize}
\item[i)]
\begin{align*}
f(G_K) \geq \frac{1}{1+ \bar \sigma} f(O),
\end{align*}
where $\bar \sigma =1- \min_{K \leq k < 2K}\frac{(1-\hat U)^{k} -(1-\hat L)^{k+1}}{\hat L}$.

\item[ii)]  \[f(G_K) \geq \frac{(1-\hat U)^{2K-1}\hat L}{\hat U} f(O).\]

\item[iii)] if $\hat L \geq 1-\frac{1}{\alpha},$
where $\alpha =\frac{1+\sqrt 5}{2}$ is the \emph{golden ratio} then\begin{align*}
f(G_K)&\geq f(O) \left(1-(1-\frac{1}{K_{ \eta}})^{K}\right) \geq \left(1-(1-\frac{1}{K_{ \bar \eta}})^{K}\right),
\end{align*}
where $K_{\eta}= ({1-\eta^{K}})/({1-\eta})$ and $\eta =\max_{a_{i}, a_{j}} \frac{(1-p^{i}(a_{i}))p^{j}(a_j) }{p^i(a_j)}$; $K_{\bar\eta}= ({1-\bar\eta^{K}})/({1-\bar\eta})$ and $\bar \eta = \frac{(1-\hat L)\hat U}{\hat L}$.
\end{itemize}
\end{Theorem}
\begin{IEEEproof}
The proofs for parts i) and ii) are omitted.
The main idea is to apply Theorem~\ref{thm4} and the calculation of the total backward curvature can be calculated in the same way as the case of uniform structure.

iii)  Now let us consider the postfix monotone property required in Theorem~\ref{thm5}: $f(G_K\oplus O) \geq f(O)$. This condition is much weaker than that in Theorem~\ref{thm2}, and can be rewritten as
\[
\prod_{j=1}^K (1-p^j(o_j)) \geq \prod_{t=1}^K (1- p^t(a_t^*)) \prod_{j=1}^{K} (1-p^{j+K}(o_j)).
\]
A sufficient condition for the above inequality is
$1-\hat U \geq (1-\hat L)^2.$
Recall that the function is string submodular if
\begin{align*}
\eta&\leq \frac{(1-\hat L)\hat U}{\hat L}\leq 1.
\end{align*} Therefore, we have $\hat U \leq 1/\alpha$ and $1-\hat U \geq (1-\hat L)^2$ holds.

\end{IEEEproof}
\subsubsection{Special Cases}
Now let us consider the special case where $p^j(a)$ is non-increasing over $j$ for each $a\in \bbA$. It is easy to show that the function is string submodular. Moreover, the elemental forward curvature is
\begin{align*}
\eta &=\max_{a_{i}, a_{j}} \frac{(1-p^{i}(a_{i}))p^{j}(a_j) }{p^i(a_j)}\\
&\leq \max_{a_i} (1-p^i(a_i))\\
&\leq 1-\hat L.
\end{align*}
Therefore, using this upper bound of the elemental forward curvature, we can provide a better approximation than $(1-e^{-1})$ for the greedy strategy for the uniform matroid case. We can also provide a good approximation for the greedy strategy for the non-uniform matroid case.

Consider the special case where $p^j(a)$ is non-decreasing over $j$ for each $a\in \bbA$. In this case, we have
\begin{align*}
\sigma(O)&\leq \hat\sigma\\
&=1-\min_{a \in \bbA, K\leq|M|< 2K}\left\{\frac{f( (a)\oplus M)-f( M)}{f( (a)) -f(\varnothing)}\right\}\\
& \leq 1-\prod_{j=1}^{|M|}(1-p^j(a_j))\\
&\leq 1-(1-\hat U)^{2K-1}.
\end{align*}
Therefore, we can provide a better approximation than $(1-e^{-1})$ for the greedy strategy using this upper bound of $\sigma(O)$ for the uniform matroid case. We can also provide a good approximation for the greedy strategy for the non-uniform matroid case.

\subsection{Maximizing the Information Gain}

In this part, we present an application of our results on string submodular functions to sequential Bayesian parameter estimation. Bayesian estimation has been studied intensively from various perspectives \cite{60s}--\nocite{70s,80s,90s,00s}\cite{10s}. This work is the first to consider the problem from the string submodularity perspective.

Consider a signal of interest $x\in \bbR^N$ with normal prior distribution $\mathcal N(\mu, P_0)$. In our example, we assume that $N=2$ for simplicity; our analysis easily generalizes to dimensions larger than 2. Let $\mathbb D$ denote the set of diagonal positive-semidefinite $2\times 2$ matrices with unit Frobenius norm:
\[
\mathbb D=\{\text{Diag}(\sqrt e,\sqrt{1-e}): e\in [0,1]\}.
\]
At each stage $i$, we choose a measurement matrix $A_i\in \mathbb D$ to get an observation $y_i$, which is corrupted by additive zero-mean Gaussian noise $\omega_i\sim \mathcal N(0, R_{\omega_i\omega_i})$:
\[
y_i=A_i x+\omega_i.
\]
Let us denote the posterior distribution of $x$ given $(y_1,y_2,\ldots, y_k)$ by $\mathcal N(x_k, P_k)$. The recursion for the posterior covariance $P_k$ is given by
\begin{align*}
P_k^{-1} &=P_{k-1}^{-1}+A_k^T R_{\omega_k\omega_k}^{-1} A_k \\
&=P_0^{-1} +\sum_{i=1}^k A_i^T R_{\omega_i\omega_i}^{-1} A_i.
\end{align*}
The entropy of the posterior distribution of $x$ given $(y_1,y_2,\ldots, y_k)$ is
$H_k=\frac{1}{2} \log \det P_k +\log (2\pi e).$
The information gain given $(A_1,A_2,\ldots, A_k)$ is
\begin{align*}
f((A_1,A_2,\ldots, A_k)) &=H_0-H_k \\
& = \frac{1}{2}(\log\det P_0 -\log\det P_k).
\end{align*}
The objective is to choose a string of measurement matrices subject to a length constraint $K$ such that the information gain is maximized.

The optimality of the greedy strategy and the measurement matrix design problem are considered in \cite{liu2012greedy} and \cite{carson2012communications}, respectively.
Suppose that the additive noise sequence is independent and identically distributed. Then, it is easy to see that
$f((A_1,A_2,\ldots, A_k))=f(\mathcal P(A_1,A_2,\ldots, A_k))$
for all permutations $\mathcal P$. Moreover, the information gain is a submodular set function and $f(\emptyset)=0$; see \cite{krause2007near}. Therefore, the greedy strategy achieves at least a $(1-e^{-1})$-approximation of the optimal strategy.

Consider the situation where the additive noise sequence is independent but \emph{not} identically distributed. Moreover, let us assume that $R_{\omega_i\omega_i}=\sigma^2_i \mathscr I$, where $\mathscr I$ denotes the identity matrix. In other words, the noise at each stage is white but the variances $\sigma_i$ depend on~$i$. The prefix monotone property is easy to see: We always gain by adding extra (noisy) measurements.

Now we investigate the sensitivity of string submodularity with respect to the varying noise variances.
\begin{Proposition} $f$ is string submodular if and only if $\sigma_i$ is monotone non-decreasing with respect $i$.
\end{Proposition}
\begin{IEEEproof}
 The sufficiency part is easy to understand: The information gain at a later stage certainly cannot be larger than the information gain at an earlier stage because the measurement $y_i$ becomes noisier as $i$ increases. We show the necessity part by contradiction. Suppose that the function is string submodular and there exists $k$ such that $\sigma_k \geq \sigma_{k+1}$. Suppose that the posterior covariance at stage $k-1$ is $\text{Diag}(s_{k-1},t_{k-1})$ and we choose $A_k=\text{Diag}(1,0)$, $A_{k+1}=\text{Diag}(0,1)$. We have
\begin{align*}
&\quad f(A_{k}\oplus A_{k+1}) -f(A_k)\\
&=\frac{1}{2}\log(1+t_{k}\sigma_{k+1}^{-2})\\
&= \frac{1}{2}\log (1+t_{k-1}\sigma_{k+1}^{-2})\\
&\geq\frac{1}{2} \log (1+t_{k-1}\sigma_{k}^{-2})\\
&=f(A_{k+1})-f(\varnothing).
\end{align*}
This contradicts the diminishing-return property and completes the argument.
\end{IEEEproof}

It is easy to show $\hat\eta \leq \eta \leq 1$ if and only if the sequence of noise variance is non-decreasing. In this case, we can compute the elemental curvature explicitly with additional information on how quickly the noise variance increases, which in turn provides performance bounds (better than $(1-e^{-1})$ for uniform matroid case and better than $1/2$ for non-uniform matroid case) for the greedy strategy.

For general cases where the noise variance sequence is not necessarily non-decreasing, we will provide an upper bound for the $K$-elemental forward curvature $\hat \eta$.
For simplicity, let $P_0=\text{Diag}(s_0,t_0)$. Without loss of generality, we assume that $s_0\geq t_0$. Let $M=(A_1,A_2,\ldots, A_{|M|})$ where $A_k=\text{Diag}(\sqrt{e_k},\sqrt{1-e_k})$ for $k=1,\ldots, |M|$. Let $
P_{|M|}=\text{Diag}(s_{|M|},t_{|M|})
$
where

\[
s_0^{-1}\leq s_{|M|}^{-1}=s_0^{-1}+\sum_{i=1}^{|M|} \sigma_i^{-2}e_i \leq s_0^{-1}+\sum_{i=1}^{|M|} \sigma_i^{-2},
\]
\[
t_0^{-1}\leq t_{|M|}^{-1}=t_0^{-1}+\sum_{i=1}^{|M|} \sigma_i^{-2}(1-e_i) \leq t_0^{-1}+\sum_{i=1}^{|M|} \sigma_i^{-2},
\]
and \[s_{|M|}^{-1}+t_{|M|}^{-1}=s_0^{-1}+t_0^{-1}+\sum_{i=1}^{|M|} \sigma_i^{-2}.\]

Next we provide an upper bound for $\hat \eta$.

\begin{Proposition}
Suppose that $\sigma_i\in [a,b]$ for each $i$, where $0<a<b$. Then, we have
\[
\hat \eta  \leq \frac{\log\frac{1}{4} (1+s_0 t_0^{-1}+2s_0{K}a^{-2})(1+\frac{s_0^{-1}+a^{-2}}{(t_0^{-1}+b^{-2})})}{\log(1+t_0(1+t_0{(2K-2)} a^{-2})^{-1} b^{-2})}.
\]

\end{Proposition}
\begin{IEEEproof}
We first derive an upper bound for the numerator in \eqref{curvature} (definition of $K$-elemental forward curvature), which is given by~\eqref{eqn:long1} on the next page.
\begin{figure*}[tbp]
\normalsize
\begin{align}
\label{eqn:long1}
\nonumber
&\quad f( M\oplus(A_i) \oplus (A_j))-f( M\oplus (A_i))=\frac{1}{2}\log (1+s_{|M|+1} \sigma_{|M|+2}^{-2} e_j)(1+t_{|M|+1}\sigma_{|M|+2}^{-2} (1-e_j))\\
\nonumber
&=\frac{1}{2}\left(\log (s_{|M|+1}^{-1}+ \sigma_{|M|+2}^{-2} e_j)(t_{|M|+1}^{-1}+\sigma_{|M|+2}^{-2} (1-e_j) )+\log s_{|M|+1}t_{|M|+1}\right)\\
\nonumber
&\leq \frac{1}{2} (\log   \left(\frac{s_{|M|+1}^{-1}+ \sigma_{|M|+2}^{-2} e_j+t_{|M|+1}^{-1}+\sigma_{|M|+2}^{-2} (1-e_j)}{2}\right)^2 \\
\nonumber
&\mbox{ }+ \max(-\log (s_0^{-1}+\sum_{i=1}^{|M|+1} \sigma_i^{-2})t_0^{-1},-\log s_0^{-1}(t_0^{-1}+\sum_{i=1}^{|M|+1} \sigma_i^{-2})))\\
& = \frac{1}{2} \left(\log \left(\frac{s_0^{-1}+t_0^{-1}+\sum_{i=1}^{|M|+2} \sigma_i^{-2}}{2}\right)^2-\log s_0^{-1}(t_0^{-1}+\sum_{i=1}^{|M|+1} \sigma_i^{-2})\right)\\
\nonumber
&= \frac{1}{2}\left(\log \left(\frac{1+s_0 t_0^{-1}+s_0\sum_{i=1}^{|M|+2} \sigma_i^{-2}}{2}\right) +\log \left(\frac{s_0^{-1}+t_0^{-1}+\sum_{i=1}^{|M|+2} \sigma_i^{-2}}{2(t_0^{-1}+\sum_{i=1}^{|M|+1} \sigma_i^{-2})}\right)\right)\\
\nonumber
&\leq\frac{1}{2}\left( \log \left(\frac{1+s_0 t_0^{-1}+s_0\sum_{i=1}^{2K} \sigma_i^{-2}}{2}\right)+ \log \left(\frac{1}{2}\left(1+\frac{s_0^{-1}+\max_{i=1,\ldots,2K} \sigma_i^{-2}}{(t_0^{-1}+\sigma_1^{-2})}\right)\right)\right).
\end{align}
\hrulefill
\vspace*{4pt}
\end{figure*}

We now derive a lower bound of the denominator in \eqref{curvature} by calculating the minimum value of the denominator over all possible $A_j$. It is easy to show that the minimum is achieved at $A_j=\text{Diag}(1,0)$ or $A_j=\text{Diag}(0,1)$:
\begin{align*}
&\quad f(M \oplus (A_j))-f(M) \\
&\geq \frac{1}{2} \min( \log (1+t_{|M|} \sigma_{|M|+1}^{-2}),\log(1+s_{|M|} \sigma_{|M|+1}^{-2}))\\
& \geq\frac{1}{2}\log(1+\min(s_{|M|} \sigma_{|M|+1}^{-2},t_{|M|} \sigma_{|M|+1}^{-2}))\\
&\geq \frac{1}{2}\log(1+(t_0^{-1}+\sum_{i=1}^{2K-2} \sigma_i^{-2})^{-1} \min_{i=1,\ldots, 2K} \sigma_i^{-2}).
\end{align*}
Therefore, we can derive an upper bound for the $K$-elemental forward curvature as follows:
\begin{align*}
&\hat \eta \leq \\
 &\frac{\log\frac{1}{4} (1+s_0 t_0^{-1}+s_0\sum_{i=1}^{2K}\sigma_i^{-2})(1+\frac{s_0^{-1}+\max_{i=1,\ldots,2K} \sigma_i^{-2}}{(t_0^{-1}+\sigma_1^{-2})})}{\log(1+(t_0^{-1}+\sum_{i=1}^{2K-2} \sigma_i^{-2})^{-1} \min_{i=1,\ldots, 2K} \sigma_i^{-2})}.
\end{align*}
Using this upper bound, we can provide an approximation bound for the greedy strategy. We note that this upper bound is not extremely tight in the sense that it does not increase significantly with $K$ only if $s_0$ or $\sigma_i^{-2}$ are sufficiently small.
By substituting either $a$ or $b$ appropriately in the inequality above, we get the upper bound for $\hat \eta$ in this proposition.

\end{IEEEproof}

With the above lower bounds for $\hat \eta$, we can use Theorem~2 to provide a bound for the greedy strategy. We have the following results.
\begin{Theorem}
Suppose that $\sigma_i\in [a,b]$ for each $i$, where $0<a<b$, and the following holds:
\[
\frac{b^{-2}}{a^{-2}-b^{-2}} \geq \frac{(2K-2)^2}{4} t_0(a^{-2}+b^{-2}) +1.
\] Then, we have
\begin{align*}
f(G_K)&\geq f(O) \left(1-(1-\frac{1}{K_{\bar \eta}})^{K}\right) \\
\end{align*}
where $K_{\hat \eta}=({1-\bar\eta^{K}})/({1-\bar \eta})$ and \[\bar \eta=\frac{\log\frac{1}{4} (1+s_0 t_0^{-1}+2s_0{K}a^{-2})(1+\frac{s_0^{-1}+a^{-2}}{(t_0^{-1}+b^{-2})})}{\log(1+t_0(1+t_0{(2K-2)} a^{-2})^{-1} b^{-2})}.
\]

\end{Theorem}

\begin{IEEEproof}
The main idea of this proof is to apply the result from Theorem~2. We have provided an upper bound for the $K$-elemental forward curvature in Proposition~4 and we will substitute this upper bound to derive bound for the greedy strategy.

Let $A^*\in \mathbb D$ be a greedy action. We will show $f((A^*)\oplus M) \geq f(M)$ for all $M$ with length $k$, where $k= K, K+1, \ldots, 2K-2$. By a mathematical induction argument, this claim leads to the sufficent condition in Theorem~2: $f(G_{i}\oplus O) \geq f(O)$ for $i=1,2,\ldots, K-1$.  Let $A^*=\text{Diag}(\sqrt{e^*},\sqrt{1-e^*})$ and $M=(A_1,\ldots, A_k)$, where $A_t=\text{Diag}(\sqrt{e_t},\sqrt{1-e_t})$ for all $t$. The inequality we need to verify can be written as
\begin{align}
\nonumber
&\log (1+s_0 (\sigma_1^{-2} e^* +\sum_{t=1}^k \sigma_{t+1}^{-2} e_t))\times \\
\nonumber
& \quad(1+t_0 (\sigma_1^{-2} (1-e^*) +\sum_{t=1}^k \sigma_{t+1}^{-2}(1- e_t)))\\
&\geq \log (1+s_0 (\sum_{t=1}^k \sigma_{t}^{-2} e_t))(1+t_0 (\sum_{t=1}^k \sigma_{t}^{-2}(1- e_t))).
\label{final}
\end{align}
We first calculate the value of $e^*$. It is easy to show that the objective function after applying $(A^*)$ achieves the maximum when \[e^*=\frac{1+\frac{t_0^{-1}-s_0^{-1}}{\sigma_1^{-1}}}{2}.\]  Because $e^*$ can only take values in $[0,1]$, in the case where ${(t_0^{-1}-s_0^{-1})}/{\sigma_1^{-1}}\geq 1$, the maximum is achieved at $e^*=1$. We will present our analysis only for this case---the analysis for the case where ${(t_0^{-1}-s_0^{-1})}/{\sigma_1^{-1}}< 1$ is similar and omitted. To show the above inequality \eqref{final}, it suffices to show that
\begin{align*}
&\log (1+s_0 \sigma_1^{-2}  +(s_0 \sum_{t=1}^k \sigma_{t+1}^{-2} e_t))(1+t_0 (\sum_{t=1}^k \sigma_{t+1}^{-2}(1- e_t)))\\
&\geq \log (1+s_0 (\sum_{t=1}^k \sigma_{t}^{-2} e_t))(1+t_0 (\sum_{t=1}^k \sigma_{t}^{-2}(1- e_t))).
\end{align*}
Removing the $\log$ on both sides of the inequality, we obtain
\begin{align*}
& (1+s_0 \sum_{t=1}^k \sigma_{t+1}^{-2} e_t) (1+t_0 \sum_{t=1}^k \sigma_{t+1}^{-2}(1- e_t))\\
&+ s_0 \sigma_1^{-2}(1+t_0 \sum_{t=1}^k \sigma_{t+1}^{-2}(1- e_t)) \\
&\geq (1+s_0 \sum_{t=1}^k \sigma_{t}^{-2} e_t) (1+t_0 \sum_{t=1}^k \sigma_{t}^{-2}(1- e_t)).
\end{align*}
Rearranging terms, we obtain \eqref{eqn:long2}, where $\bbI_{t}=1$ if $\sigma_{t+1}^{-2} \leq \sigma_t^{-2}$ and $\bbI_{t}=0$ if $\sigma_{t+1}^{-2} > \sigma_t^{-2}$.
\begin{figure*}[tbp]
\normalsize
\begin{align}
\label{eqn:long2}
\nonumber
&\quad s_0 \sum_{t=1}^{k} e_t (\sigma_{t+1}^{-2} -\sigma_t^{-2}) +t_0 \sum_{t=1}^k (1-e_t) (\sigma_{t+1}^{-2} -\sigma_t^{-2})+ s_0 \sigma_1^{-2}(1+t_0 \sum_{t=1}^k \sigma_{t+1}^{-2}(1- e_t)) \\
\nonumber
&\quad+s_0t_0(\sum_{t=1}^k \sigma_{t+1}^{-2} e_t)(\sum_{t=1}^k \sigma_{t+1}^{-2} (1-e_t))-s_0t_0(\sum_{t=1}^k \sigma_{t}^{-2} e_t)(\sum_{t=1}^k \sigma_{t}^{-2} (1-e_t)) \\
& \geq s_0 \sum_{t=1}^k (\sigma_{t+1}^{-2} -\sigma_t^{-2})\bbI_{t}+t_0  \sum_{t=1}^k (\sigma_{t+1}^{-2} -\sigma_t^{-2})(1-\bbI_{t}) \\
\nonumber
&\quad+ s_0 \sigma_1^{-2}(1+t_0 \sum_{t=1}^k \sigma_{t+1}^{-2}(1- e_t))  + s_0t_0 (b^{-4}-a^{-4}) (\sum_{t=1}^k e_t)(\sum_{t=1}^k (1-e_t))\\
\nonumber
&\geq  s_0 (b^{-2}-a^{-2})+s_0 b^{-2}+ \frac{k^2}{4} s_0t_0 (b^{-4}-a^{-4}) \geq 0.
\end{align}
\hrulefill
\vspace*{4pt}
\end{figure*}

From this we obtain a sufficient condition for $f((A^*)\oplus M) \geq f(M)$ to hold:
\[
\frac{b^{-2}}{a^{-2}-b^{-2}} \geq \frac{k^2}{4} t_0(a^{-2}+b^{-2}) +1.
\]
The term on the right is monotone increasing with respect to $k$ and achieves its maximum at $2K-2$. The proof is completed.
\end{IEEEproof}

\section{Conclusion}
In this paper, we have introduced the notion of total forward/backward and elemental forward curvature for functions defined on strings. We have derived several variants of lower performance bounds, in terms of these curvature values, for the greedy strategy with respect to the optimal strategy.
Our results contribute significantly to our understanding of the underlying algebraic structure of string submodular functions. Moreover, we have investigated two applications of string submodular functions with curvature constraints.

\appendices

\section{Proof of Proposition~1}
(i) For any $M,N\in \bbA^*$ and $|M|\leq K$, $|N|\leq K$, we have shown in the proof of Theorem 1 that, there exists $a\in \bbA$ such that
\begin{align*}
&\quad f(M\oplus N)-f(M) \leq K_{\hat\eta} ( f(M\oplus (a))-f(M)).
\end{align*}

Now let us consider the optimization problem \eqref{eqn:1} with length constraint $K$. Using the property of the greedy strategy and the monotone property, we have
\begin{align*}
f(G_i)-f(G_{i-1}) &\geq   \frac{1}{K_{\hat\eta}}  (f(G_{i-1}\oplus O)-f(G_{i-1}))\\
&\geq \frac{1}{K_{\hat\eta}} (f(O)-\sigma(O)f(G_{i-1})).
\end{align*}
Therefore, by recursion, we have
\begin{align*}
f(G_K)&\geq \frac{1}{K_{\hat\eta}} f(O)+(1-\frac{\sigma(O)}{K_{\hat\eta}})f(G_{K-1}) \\
&\geq \frac{1}{K_{\hat\eta}}f(O)\sum_{i=0}^{K-1}  (1-\frac{\sigma(O)}{K_{\hat\eta}})^i \\
&=\frac{1}{\sigma(O)}\left(1-(1-\frac{\sigma(O)}{K_{\hat\eta}})^{K}\right)f(O).
\end{align*}
The second inequality simply follows from the facts that $\frac{1}{\sigma(O)}\left(1-(1-\frac{\sigma(O)}{K_{\hat\eta}})^{K}\right)$ is a monotone decreasing function of $\hat \eta$ and $\hat \eta \leq \eta$ by definition.

(ii) Using a similar argument as part (i), we have
\begin{align*}
&f(G_i)-f(G_{i-1}) \\
&\geq  \frac{1}{K_{\hat\eta}} (f(G_{i-1}\oplus O)-f(G_{i-1}))\\
&\geq \frac{1}{K_{\hat\eta}} (f(G_{i-1})-f(G_{i-1})+(1-\epsilon(G_{i-1}))f(O)).
\end{align*}
Therefore, by recursion,
\begin{align*}
f(G_K)&=\sum_{i=1}^K (f(G_i)-f(G_{i-1}) )\\
&\geq\sum_{i=1}^K \frac{1}{K_{\hat \eta}} (1-\epsilon(G_{i-1}))f(O) \\
&\geq  \frac{K}{K_{\hat\eta}}  (1-\max_{i=1,\ldots, K-1} \epsilon(G_i)) f(O).
\end{align*}
The second inequality simply follows from the facts that $\frac{K}{K_{\hat\eta}}$ is a monotone decreasing function of $\hat \eta$  and $\hat \eta \leq \eta$ by definition.

\section{Proof of Proposition~2}
(i) Using the definition of total backward curvature, we have
\[f(G_K\oplus O)-f(O) \geq (1-\sigma(O))f(G_K),\]
which implies that
\[f(G_K\oplus O)-f(G_K) \geq f(O)-\sigma(O)f(G_K).\]
Using a similar argument as that of Theorem~5, we know that
\[
f(G_K\oplus O)-f(G_K) \leq h(\hat\eta) f(G_K).
\]
Therefore, we have
\[
f(G_K) \geq \frac{1}{h(\hat\eta) +\sigma(O)}f(O).
\]
The second inequality follows from $h(\hat\eta) \leq h(\eta)$.

(ii) Using the definition of total forward curvature, we have
\[f(G_K\oplus O)-f(G_K) \geq (1-\epsilon(G_K))f(O).\]
Using a similar argument as that of Theorem~5, we know that
$f(G_K\oplus O)-f(G_K) \leq h(\hat\eta) f(G_K).$
Therefore, we have
\[
f(G_K) \geq \frac{1-\epsilon(G_K)}{h(\hat\eta)}f(O).
\]
The second inequality follows from $h(\hat\eta) \leq h(\eta)$.

\section*{Acknowledgment}

The authors wish to thank the anonymous reviewers for the careful reading of the manuscript and constructive comments that have improved the presentation.
The authors also gratefully thank Zengfu Wang for sending them a preprint of \cite{wang2012}. Finally, the authors thank Yajing Liu and Yuan Wang for their careful reading of the manuscript and many constructive comments.

\ifCLASSOPTIONcaptionsoff
  \newpage
\fi

\bibliographystyle{IEEEbib}

\begin{thebibliography}{10}



\bibitem{zhang2012submodularity}
Z.~Zhang, E.~K.~P. Chong, A.~Pezeshki, W.~Moran, and S.~D. Howard,
  ``Submodularity and optimality of fusion rules in balanced binary relay
  trees,'' in \emph{Proc. IEEE 51st Annual Conference on Decision and Control},
  Maui, HI, Dec. 2012, pp. 3802--3807.

\bibitem{2013submodularity}
Z.~Zhang, Z.~Wang, E.~K.~P. Chong, A.~Pezeshki, and W.~Moran,
  ``Near optimality of greedy strategies for string submodular functions with forward and backward curvature constraints,'' in \emph{Proc. IEEE 52nd Annual Conference on Decision and Control},
  Florence, Italy, December 10--13, 2013, pp. 5156--5161.




\bibitem{Bertsekas2000}
D.~P. Bertsekas, \emph{Dynamic programming and optimal control}, Athena Scientific, 2000.

\bibitem{cover1991universal}
T.~M. Cover, ``Universal portfolios,'' \emph{Mathematical Finance}, vol.~1,
  no.~1, pp. 1--29, Jan. 1991.

\bibitem{chong2009partially}
E.~K.~P. Chong, C.~M. Kreucher, and A.~O. Hero~III, ``Partially observable
  markov decision process approximations for adaptive sensing,'' \emph{Discrete
  Event Dynamic Systems}, vol.~19, no.~3, pp. 377--422, Sep. 2009.


\bibitem{li2014suboptimal}
C. Li and N. Elia, ``A suboptimal sensor scheduling strategy using convex optimization,'' in \emph{Proc. of American Control Conference (ACC), 2011 , pp. 3603--3608, June 29--July 1 2011}.

\bibitem{li2014stochastic}
C. Li and N. Elia, ``Stochastic sensor scheduling via distributed convex optimization,'' to appear in \emph{Automatica} 2014. 
\bibitem{mossel2007submodularity}
E.~Mossel and S.~Roch, ``On the submodularity of influence in social
  networks,'' in \emph{Proc. 39th Annual ACM Symposium on Theory of Computing},
  San Diego, California, USA, Jun. 2007, pp. 128--134.


\bibitem{tac11}
A~Krause, R.~Rajagopal, A.~Gupta, and C.~Guestrin, ``Simultaneous optimization of sensor placements and balanced schedules,''  \emph{IEEE Transactions on Automatic Control}, vol.~56, no.~10, pp. 2390--2405, Oct. 2011.


\bibitem{tutte1965lectures}
W.~Tutte, ``Lectures on matroids,'' \emph{Journal of Research of the National Bureau of Standards Section B},
  vol.~69, no. 468, pp. 1--47, Jan--Jun. 1965.

\bibitem{nemhauser1978analysis}
G.~L. Nemhauser, L.~A. Wolsey, and M.~L. Fisher, ``An analysis of
  approximations for maximizing submodular set functions---i,''
  \emph{Mathematical Programming}, vol.~14, no.~1, pp. 265--294, 1978.

\bibitem{fisher1978analysis}
M.~L. Fisher, G.~L. Nemhauser, and L.~A. Wolsey, ``An analysis of
  approximations for maximizing submodular set functions---ii,'' \emph{Mathematical Programming}, vol.~8, pp.
  73--87, 1978.

\bibitem{conforti1984submodular}
M.~Conforti and G.~Cornuejols, ``Submodular set functions, matroids and the
  greedy algorithm: tight worst-case bounds and some generalizations of the
  rado-edmonds theorem,'' \emph{Discrete Applied Mathematics}, vol.~7, no.~3,
  pp. 251--274, 1984.

\bibitem{vondrak2010submodularity}
J.~Vondr{\'a}k, ``Submodularity and curvature: the optimal algorithm,''
  \emph{RIMS Kokyuroku Bessatsu B}, vol.~23, pp. 253--266, 2010.

\bibitem{wang2012}
Z.~Wang, W.~Moran, X.~Wang, and Q.~Pan, ``Approximation for maximizing monotone
  non-decreasing set functions with a greedy method,'' \emph{Journal of Combinatorial Optimization}, DOI: 10.1007/s10878-014-9707-3, Jan. 2014.

\bibitem{streeter2008online}
M.~Streeter and D.~Golovin, ``An online algorithm for maximizing submodular
  functions,'' in \emph{Proc. 22nd Annual Conference on Neural Information
  Processing Systems}, Vancouver, B.C., Canada, Dec. 2008, pp. 1577--1584..



\bibitem{golovin2011adaptive}
D.~Golovin and A.~Krause, ``Adaptive submodularity: Theory and applications in
  active learning and stochastic optimization,'' \emph{Journal of Artificial
  Intelligence Research}, vol.~42, no.~1, pp. 427--486, Sep. 2011.


\bibitem{alaei2010maximizing}
S.~Alaei and A.~Malekian, ``Maximizing sequence-submodular functions and its
  application to online advertising,'' \emph{arXiv preprint arXiv:1009.4153},
  2010.


\bibitem{buchbinder2012tight}
N.~Buchbinder, M.~Feldman, J.~S. Naor, and R.~Schwartz, ``A tight linear time
  (1/2)-approximation for unconstrained submodular maximization,'' in
  \emph{Proc. IEEE 53rd Annual Symposium on Foundations of Computer Science},
  New Brunswick, NJ, Oct. 2012, pp. 649--658.

\bibitem{calinescu2011maximizing}
G.~Calinescu, C.~Chekuri, M.~P{\'a}l, and J.~Vondr{\'a}k, ``Maximizing a
  monotone submodular function subject to a matroid constraint,'' \emph{SIAM
  Journal on Computing}, vol.~40, no.~6, pp. 1740--1766, Nov. 2011.

\bibitem{chakrabarty2010approximability}
D.~Chakrabarty and G.~Goel, ``On the approximability of budgeted allocations
  and improved lower bounds for submodular welfare maximization and gap,''
  \emph{SIAM Journal on Computing}, vol.~39, no.~6, pp. 2189--2211, Mar. 2010.

\bibitem{vondrak2011submodular}
J.~Vondr{\'a}k, C.~Chekuri, and R.~Zenklusen, ``Submodular function
  maximization via the multilinear relaxation and contention resolution
  schemes,'' in \emph{Proc. 43rd Annual ACM Symposium on Theory of Computing}, San Jose, California, USA,
 Jun. 2011, pp. 783--792.

\bibitem{dobzinski2006improved}
S.~Dobzinski and M.~Schapira, ``An improved approximation algorithm for
  combinatorial auctions with submodular bidders,'' in \emph{Proc. 17th
  ACM-SIAM Symposium on Discrete Algorithm}, 2006, pp. 1064--1073.

\bibitem{feige1998threshold}
U.~Feige, ``A threshold of $\ln n$ for approximating set cover,'' \emph{Journal
  of the ACM}, vol.~45, no.~4, pp. 634--652, Jul. 1998.

\bibitem{feige2006approximation}
U.~Feige and J.~Vondrak, ``Approximation algorithms for allocation problems:
  Improving the factor of $1-1/e$,'' in \emph{Proc. 47th IEEE Symposium on
  Foundations of Computer Science}, Berkeley, CA,
Oct. 2006, pp. 667--676.

\bibitem{feige2010submodular}
U.~Feige and J.~Vondr{\'a}k, ``The submodular welfare problem with demand
  queries,'' \emph{Theory of Computing}, vol.~6, pp. 247--290, Aug. 2010.

\bibitem{filmus2012tight}
Y.~Filmus and J.~Ward, ``A tight combinatorial algorithm for submodular
  maximization subject to a matroid constraint,'' in \emph{Proc. IEEE 53rd
  Annual Symposium on Foundations of Computer Science}, New Brunswick, NJ,
  2012, pp. 659--668.

\bibitem{kulik2009maximizing}
A.~Kulik, H.~Shachnai, and T.~Tamir, ``Maximizing submodular set functions
  subject to multiple linear constraints,'' in \emph{Proc. 20th ACM-SIAM
  Symposium on Discrete Algorithms}, New York, New York, Jan. 2009, pp. 545--554.

\bibitem{lee2009non}
J.~Lee, V.~S. Mirrokni, V.~Nagarajan, and M.~Sviridenko, ``Non-monotone
  submodular maximization under matroid and knapsack constraints,'' in
  \emph{Proc. ACM Symposium on Theory of Computing}, Bethesda, MD, May--Jun. 2009, pp. 323--332.

\bibitem{lee2010submodular}
J.~Lee, M.~Sviridenko, and J.~Vondr{\'a}k, ``Submodular maximization over
  multiple matroids via generalized exchange properties,'' \emph{Mathematics of
  Operations Research}, vol.~35, no.~4, pp. 795--806, Nov. 2010.

\bibitem{nemhauser1978best}
G.~L. Nemhauser and L.~A. Wolsey, ``Best algorithms for approximating the
  maximum of a submodular set function,'' \emph{Mathematics of Operations
  Research}, vol.~3, no.~3, pp. 177--188, 1978.

\bibitem{sviridenko2004note}
M.~Sviridenko, ``A note on maximizing a submodular set function subject to a
  knapsack constraint,'' \emph{Operations Research Letters}, vol.~32, no.~1,
  pp. 41--43, Jan. 2004.

\bibitem{vondrak2008optimal}
J.~Vondr{\'a}k, ``Optimal approximation for the submodular welfare problem in the
  value oracle model,'' in \emph{Proc. 40th ACM Symposium on Theory of
  Computing}, Victoria, British Columbia, Canada, May. 2008, pp. 67--74.

\bibitem{shamaiah2010greedy}
M.~Shamaiah, S.~Banerjee, and H.~Vikalo, ``Greedy sensor selection: Leveraging
  submodularity,'' in \emph{Proc. 49th IEEE Conference onDecision and Control}, Atlanta, GA,
Dec. 2010, pp. 2572--2577.

\bibitem{ageev2004pipage}
A.~A. Ageev and M.~I. Sviridenko, ``Pipage rounding: A new method of
  constructing algorithms with proven performance guarantee,'' \emph{Journal of
  Combinatorial Optimization}, vol.~8, no.~3, pp. 307--328, 2004.


\bibitem{nesstac}
Z.~Zheng and N.~B.~Shroff, ``Sumbodular utility maximization for deadline constrained data collection in sensor networks,''
\emph{IEEE Transactions on Automatic Control}, to appear, DOI: 10.1109/TAC.2014.2321683.



\bibitem{feige2004approximating}
U.~Feige, L.~Lov{\'a}sz, and P.~Tetali, ``Approximating min sum set cover,''
  \emph{Algorithmica}, vol.~40, no.~4, pp. 219--234, 2004.

\bibitem{munagala2005pipelined}
K.~Munagala, S.~Babu, R.~Motwani, and J.~Widom, ``The pipelined set cover
  problem,'' in \emph{Proc. 10th International Conference on Database Theory}, Edinburgh, UK, Jan. 2005, pp. 83--98.

\bibitem{kempe2005influential}
D.~Kempe, J.~Kleinberg, and {\'E}.~Tardos, ``Influential nodes in a diffusion
  model for social networks,'' in \emph{Proc. 32nd International Conference on
  Automata, Languages and Programming}, Lisbon, Portugal, Jul. 2005, pp. 1127--1138.

\bibitem{lu2003coverage}
J.~Lu and T.~Suda, ``Coverage-aware self-scheduling in sensor networks,'' in
  \emph{Proc. IEEE 18th Annual Workshop on Computer Communications}, Dana Point, California, Oc. 2003, pp.
  117--123.

\bibitem{tacScheduling}
F. Iannello and O. Simeone,
``On the optimal scheduling of independent, symmetric and time-sensitive tasks,''
\emph{IEEE Transactions on Automatic Control},
vol.~58, no.~9, pp.~2421--2425 2013.


\bibitem{60s} Y. C. Ho and R. Lee, ``A Bayesian approach to problems in stochastic estimation and control,''
\emph{IEEE Transactions on Automatic Control},
vol.~9, no.~4, pp.~333--339, 1964.
\bibitem{70s} D. L. Alspach and H. W. Sorenson,
``Nonlinear Bayesian estimation using Gaussian sum approximations,''
\emph{IEEE Transactions on Automatic Control},
vol.~17, no.~4, pp.~439--448, 1972.
\bibitem{80s} S. C. Kramer and H. W. Sorenson,
``Bayesian parameter estimation,''
\emph{IEEE Transactions on Automatic Control},
vol.~33, no.~2, pp.~217--222, 1988.
\bibitem{90s} J. C. Spall and S. D. Hill,
``Least-informative Bayesian prior distributions for finite samples based on information theory,''
\emph{IEEE Transactions on Automatic Control},
vol.~35, no.~5, pp.~580--583, 1990.
\bibitem{00s} B. Ait-El-Fquih and F. Desbouvries,
``On Bayesian fixed-interval smoothing algorithms,''
\emph{IEEE Transactions on Automatic Control},
vol.~53, no.~10, pp.~2437--2442, 2008.
\bibitem{10s} G. Terejanu, P. Singla, T. Singh, and P. D. Scott,
``Adaptive Gaussian sum filter for nonlinear Bayesian estimation,''
\emph{IEEE Transactions on Automatic Control},
vol.~56, no.~9, pp.~2151--2156, 2011.



\bibitem{liu2012greedy}
E.~Liu, E.~K.~P. Chong, and L.~L. Scharf, ``Greedy adaptive measurements with
  signal and measurement noise,'' \emph{IEEE Transactions on Information Theory}, vol. 60, no. 4, pp. 2269--2280, Apr. 2014.

\bibitem{carson2012communications}
W.~R. Carson, M.~Chen, M.~R. Rodrigues, R.~Calderbank, and L.~Carin,
  ``Communications-inspired projection design with application to compressive
  sensing,'' \emph{SIAM Journal on Imaging Sciences}, vol.~5, no.~4, pp.
  1185--1212, 2012.

\bibitem{krause2007near}
A.~Krause and C.~Guestrin, ``Near-optimal observation selection using
  submodular functions,'' in \emph{Proc. National Conference on Artificial
  Intelligence}, vol.~22, no.~2, Vancouver, British Columbia, Canada, Jul. 2007, pp. 1650--1654.

\end{thebibliography}

\end{document}